\let\TeXyear\year
\documentclass{ieeeaccess}
\let\setyear\year
\let\year\TeXyear

\setyear{2023}
\usepackage{cite}
\usepackage{amsmath,amssymb,amsfonts}
\usepackage{algorithmic}
\usepackage{graphicx}
\usepackage{textcomp}
\usepackage[colorlinks,urlcolor=blue]{hyperref}
\hypersetup{
    colorlinks=true,
    linkcolor=blue,
    citecolor=blue,
    urlcolor=blue,
    }
\usepackage{xfrac}   
\def\BibTeX{{\rm B\kern-.05em{\sc i\kern-.025em b}\kern-.08em
    T\kern-.1667em\lower.7ex\hbox{E}\kern-.125emX}}
\usepackage[english]{babel}

\usepackage{subcaption}
\usepackage[labelsep=period]{caption}
\usepackage{placeins}

\usepackage[exponent-product=\cdot]{siunitx}

\usepackage[nameinlink,capitalize]{cleveref}
\Crefname{figure}{Fig.}{Figs.}
\Crefname{table}{Table}{Tables}
\DeclareCaptionFormat{figcapfont}{\figcapfont \textbf{\textcolor{accessblue}{#1#2}}\textsf{#3}}
\def\subfigcapfont{\rmfamily\fontencoding{T1}\fontseries{n}\fontsize{8}{9.6}\selectfont} 
\DeclareCaptionFormat{subfigcapfont}{\subfigcapfont #1#2#3}
\renewcommand{\thetable}{\Roman{table}}

\usepackage{float}

\usepackage{tabularx}
\usepackage{booktabs}

\newcommand{\qd}[1]{quantum dot#1}

\newcommand{\sd}[1]{sensor dot#1}
\newcommand{\Sd}[1]{Sensor dot#1}
\newcommand{\nest}[1]{noise estimation#1}

\newcommand{\np}[1]{\textit{NumPy#1}}
\usepackage[nolist,nohyperlinks]{acronym}
\begin{acronym}[\(CQ_{0.99}R\)]
	\acro{awgn}[AWGN]{additive white Gaussian noise} 
	\acro{dct}[DCT]{discrete cosine transformation}
	\acro{iqr}[IQR]{interquartile range}
	\acro{msg}[MSG]{minimal signal gradient}
	\acro{pca}[PCA]{principal component analysis}
	\acro{pcb}[PCB]{printed circuit board}
	\acro{rf}[RF]{radio frequency}
	\acro{rmse}[RMSE]{root mean square error}
	\acro{roi}[FOI]{flank-of-interest}
	\acro{std}[STD]{standard deviation}
	\acro{qc}[QC]{quantum computing}
	\acro{gm}[\(G_m\)]{measured signal}
	\acro{gm1d}[\(G_{m, 1D}\)]{one-dimensional measurement signal}
	\acro{gsd}[\(G_{sd}\)]{signal at the sensor dot}
	\acro{n}[\(n\)]{noise}
	\acro{nind}[\(n_{const}\)]{constant noise}
	\acro{ngrad}[\(n_{grad}\)]{gradient-dependent noise}
	\acro{nsig}[\(n_{lin}\)]{linearly signal-dependent noise}
	\acro{ndist}[\(\mathcal{N}\)]{normal distribution}
	\acro{sigma}[\(\sigma{}\)]{standard deviation}
	\acro{xi}[\(X\)]{normally distributed random variable}
	\acro{cq95}[\(CQ_{0.95}\)]{centralized 95\%{}-quantile}
	\acro{cq99}[\(CQ_{0.99}\)]{centralized 99\%{}-quantile}
	\acro{cq99r}[\(C_{0.99}R\)]{centralized 99\%{}-quantile range}
\end{acronym}

\usepackage{pifont}
\usepackage[final]{microtype}

\begin{document}
    \doi{-}
    
    \title{On Noise-Sensitive Automatic Tuning of Gate-Defined Sensor Dots}
    
    \author{\uppercase{Fabian~Hader}\authorrefmark{1},
            \uppercase{Jan~Vogelbruch}\authorrefmark{1},
            \uppercase{Simon~Humpohl}\authorrefmark{2},
            \uppercase{Tobias~Hangleiter}\authorrefmark{2},
            \uppercase{Chimezie~Eguzo}\authorrefmark{1},
            \uppercase{Stefan~Heinen}\authorrefmark{1},
            \uppercase{Stefanie~Meyer}\authorrefmark{1},
            \uppercase{and Stefan~van~Waasen}\authorrefmark{1,3}}
    \address[1]{Central Institute of Engineering, Electronics and Analytics
    ZEA-2 -- Electronic Systems, Forschungszentrum Jülich GmbH, 52425 Jülich,
    Germany}
    \address[2]{JARA-FIT Institute for Quantum Information, Forschungszentrum Jülich GmbH and RWTH Aachen University, 52074 Aachen, Germany}
    \address[3]{Faculty of Engineering -- Communication Systems, University of Duisburg-Essen, 47057 Duisburg, Germany}
    \tfootnote{This work was supported by the Impulse and Networking Fund of the Helmholtz Association.}

    \markboth
    {Hader \headeretal: On Noise-Sensitive Automatic Tuning of Gate-Defined Sensor Dots}
    {Hader \headeretal: On Noise-Sensitive Automatic Tuning of Gate-Defined Sensor Dots}

    \corresp{Corresponding author: Fabian Hader (email: f.hader@fz-juelich.de).}

    \begin{abstract}
    	In gate-defined \qd{} systems, the conductance change of electrostatically coupled \sd{s} allows the observation of the quantum dots' charge and spin states. Therefore, the \sd{} must be optimally sensitive to changes in its electrostatic environment. A series of conductance measurements varying the two sensor-dot-forming barrier gate voltages serve to tune the dot into a corresponding operating regime. In this paper, we analyze the noise characteristics of the measured data and define a criterion to identify continuous regions with a sufficient signal-gradient-to-noise ratio. Hence, accurate noise estimation is required when identifying the optimal operating regime. Therefore, we evaluate several existing noise estimators, modify them for 1D data, optimize their parameters, and analyze their quality based on simulated data. The estimator of Chen et al. turns out to be best suited for our application concerning minimally scattering results. Furthermore, using this estimator in an algorithm for flank-of-interest classification in measured data shows the relevance and applicability of our approach.
    \end{abstract}
    

    \titlepgskip=-15pt
    \maketitle

    \section{Introduction}
    \label{sec:introduction}
        \PARstart{Q}{uantum} dots are a promising scalable platform for quantum computations. However, the isolation of electrons in quantum dots is a non-trivial task that needs to be automated. One can observe the charge and spin states in gate-defined \qd{s} by the conductance change of a nearby electrostatically coupled \sd{}. It is optimally sensitive if the \sd{} gate voltages reach a particular operating regime during the tuning. To locate the operating point, we carry out reflectometry measurements\footnote{Although in our work reflectometry measurements are performed, the results should also be transferable to other measurement techniques (e.g. DC)} of the conductance through the \sd{} while varying the barrier gate voltages, \(SL\) and \(SR\), forming the \sd{}.
	    This measured signal, proportional to the conductance \(G_m(SL, SR)\), comprises the undisturbed \sd{} signal overlayed with some noise. Thus, as part of the autotuning development, our primary focus is to identify continuous regions with sufficient signal-gradient-to-noise ratio.
	    
\begin{figure}[!tbp]
    	\centering
		\includegraphics[width=1\linewidth]{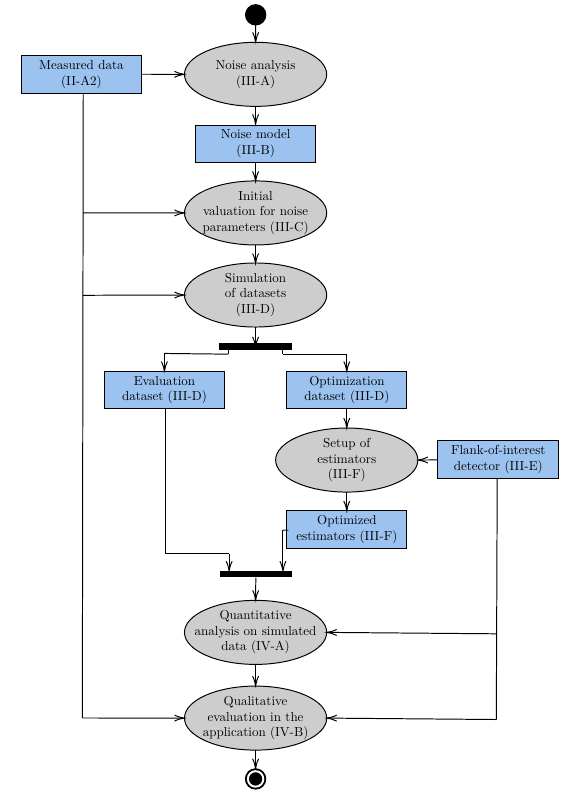}
            \caption{Activity diagram of the methodological procedure. The numbers in brackets refer to the associated section.}
            \label{fig:methodology}
        \end{figure}

	    \cref{fig:methodology} illustrates the methodological procedure of our investigation. Section \ref{ssec:tuning} covers the tuning procedure and the requirements for suitable voltage regions. The noise characteristics must be accurately estimated when identifying the optimal operating regime. Many noise estimators exist for 2D data applications (see \cref{ssec:nest}). However, as our \sd{} data sampling is highly anisotropic, the estimation can only be performed on 1D data. Thus, only estimators that are adaptable to 1D come into consideration. Furthermore, a quantitative analysis of the estimator results requires a known ground truth. Therefore, an inevitable first step is to analyze the noise sources and generate a theoretical model. Then, generated realistic sample data, composed of simulated noise and approximated clean \sd{} data, constitute the ground truth. \cref{ssec:noise_analysis,ssec:noise_model,ssec:initial_guess_noise,ssec:eval_data} cover the noise analysis and generation of data.

	 After quantitatively comparing the estimators with simulated data in \cref{ssec:results_artificial}, the qualitative evaluation in \cref{ssec:results_real} analyzes the impact of the estimators' characteristics on the detection of qualified regions in measured data.

    \newpage
     
    \section{Background}
    \label{sec:background}
    
    \subsection{Sensor dot tuning}
    \label{ssec:tuning}
    
    Semiconductor gate-defined \qd{} systems often incorporate charge sensors to measure the quantum dot's charge state which allows accessing dot parameters like capacitive coupling and tunnel coupling. Moreover, the electron spin which is typically used to encode and process quantum information can be read out via spin-to-charge conversion\cite{hanson_2007b}.
    
    One way to realize a charge sensor is a capacitively coupled \qd{} called a \sd{}. In the correct operating regime, changes in the local electrostatic environment directly affect the conductance through the \sd{} \cite{smith_2017}, thus providing measurement information about the \qd{}'s charge state. The key to tuning a \sd{} is to observe the energy of the dot via its conductance characteristics as a function of the applied gate voltages \cite{vanderwiel_2002a}.
    
\begin{figure}[!tbp]
    	\centering
		\includegraphics[width=0.9\linewidth]{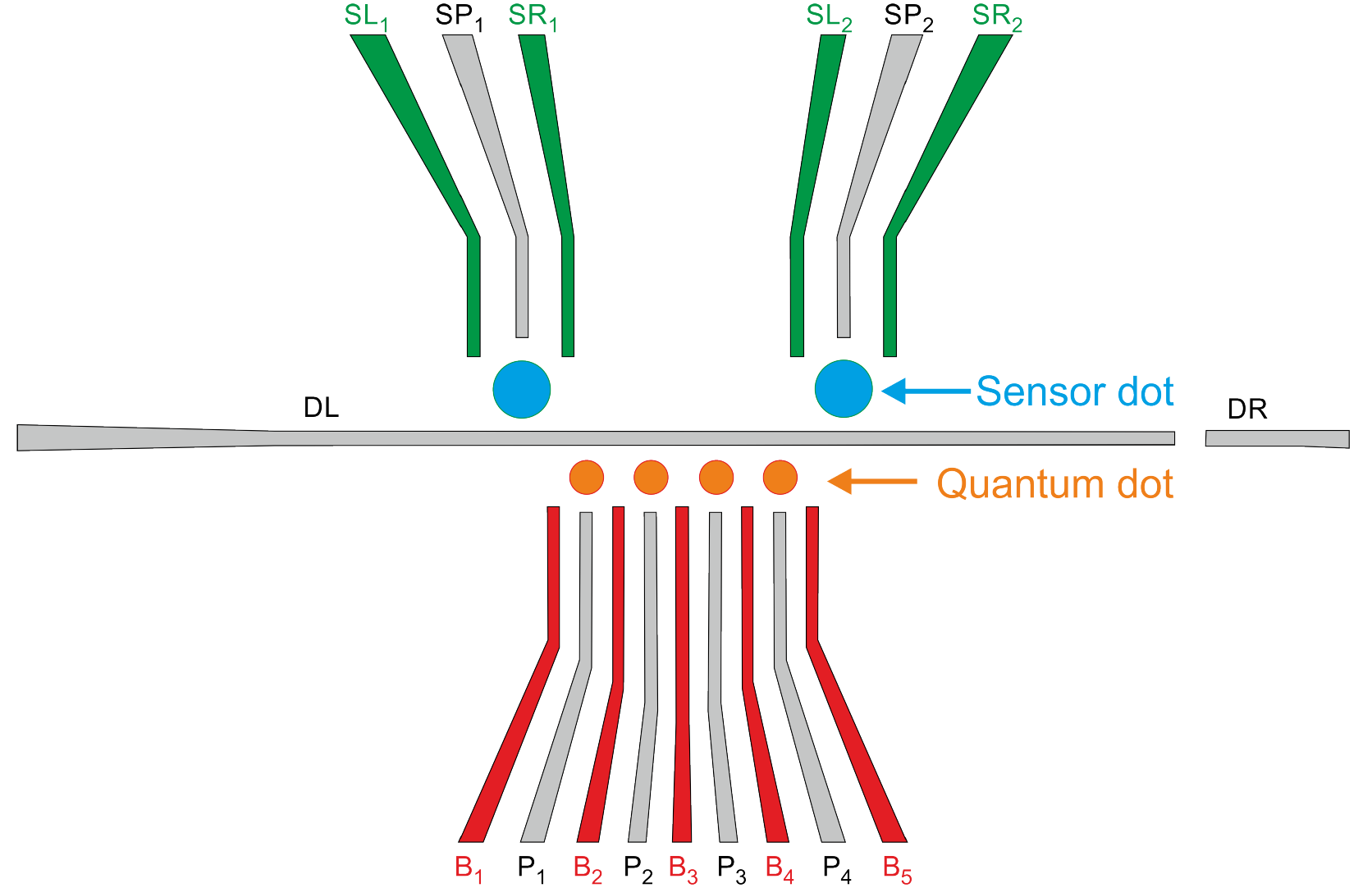}
    	\caption[Arrangement of the \sd{} gates]{Arrangement of the \sd{} gates. Blue spots mark \sd{} regions and orange spots \qd{} areas. The gates \(SL_1\), \(SP_1\), and \(SR_1\) form the left \sd{}, the gates \(SL_2\), \(SP_2\), and \(SR_2\) the right one.}
    	\label{fig:sd_gates}
    \end{figure}

\begin{figure*}
    	\centering
		\includegraphics[width=1\linewidth]{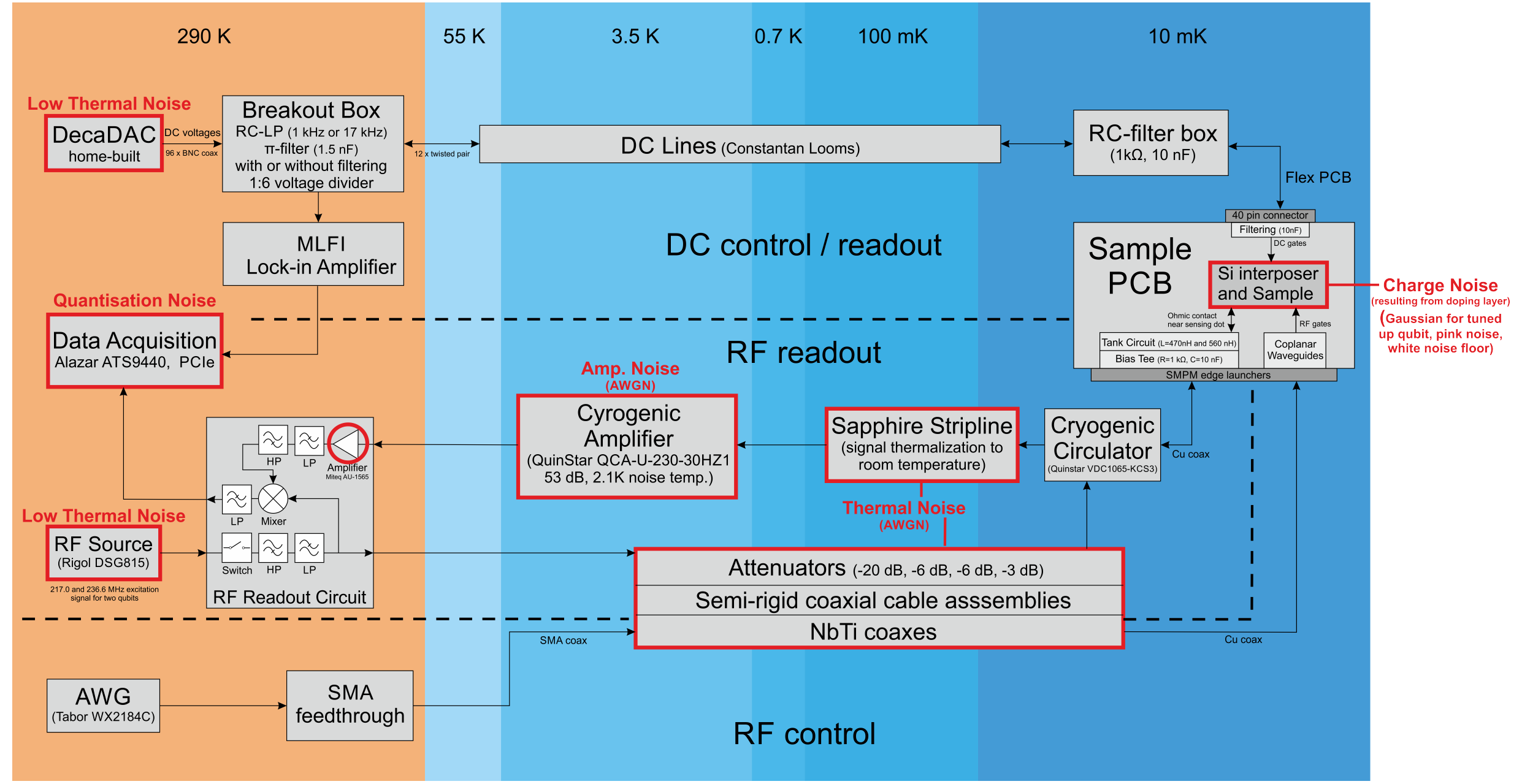}
    	\caption[Signal path]{Diagram of the signal path. The sample \ac{pcb} contains the current sample with the gates for the \sd{s} and \qd{s}. In the DC part, a digital-to-analog converter (DecaDAC) processes and generates the gate voltages. Then, the breakout box filters these voltages before arriving at the area cooled to 55 Kelvin. Here, 12 twisted pairs of DC lines (constantan looms) transmit the signals to the RC filter box, which operates at only 10 mK. After filtering, the voltages reach the gates on the sample \ac{pcb}. 
    		The \ac{rf} part implements a homodyne detection setup. A single tone generated by an \ac{rf} source is sent via a coaxial cable and a circulator to the sample \ac{pcb} where it is matched to the sensor dot impedance by an LC tank circuit. The circulator directs the signal reflected at the sample to a second coaxial cable. This signal is thermalized by a sapphire stripline and amplified at 3.5 K and at room temperature, where it is demodulated, low-pass filtered, and finally digitized using a PCI Express card.
    	}
    	\label{fig:signal_path}
    \end{figure*}

    For the present experiment, \cref{fig:sd_gates} illustrates the arrangement of the gates that form the regions for the sensor and the quantum dots. \cite{botzem_2017, cerfontaine_2019} describe the overall experiment setup, and \cref{fig:signal_path} shows the schematic representation of the signal path.
    
    The works mentioned report on two types of existent noise that are relevant for this work: charge noise and thermal noise (see \cref{tab:sources_summary}). According to \cite{dial_2013} charge noise shows pink characteristics. 
    For pink noise, also known as \(\sfrac{1}{f}\)-noise, the power spectral density is proportional to $f^{-\alpha}$ with typically $0.5 \lesssim \alpha \lesssim 2$\cite{Paladino2014,connors2022,Struck2019}. On the other hand, thermal noise, also known as Johnson-Nyquist noise, is approximately white and has a nearly Gaussian amplitude distribution if limited to a finite bandwidth \cite{barry_2004}. Therein, white denotes a constant noise power spectral density.
    
    \begin{table}[!tbp]
    	\centering
    	\noindent
    	\caption{Most relevant noise sources in quantum dot systems.}
    	\label{tab:sources_summary}
    	\begin{minipage}{\linewidth}
    		\begin{tabularx}{\linewidth}{lXlc}
    			\toprule
    			\textbf{Noise type} & \textbf{Noise source} & \textbf{Characteristics} \\
    			\midrule
    			Charge noise & Fluctuating charges in the heterostructure & Pink, Gaussian \\
    			\midrule
    			Johnson-Nyquist noise & Thermal agitation of the charge carriers in the amplifier & White, Gaussian \\
    			\bottomrule
    		\end{tabularx}
    	\end{minipage}
    \end{table}
    
    \subsubsection{\Sd{} preparation}
    \label{sssec:device_preparation}
    
\begin{figure}
		\includegraphics[width=\linewidth]{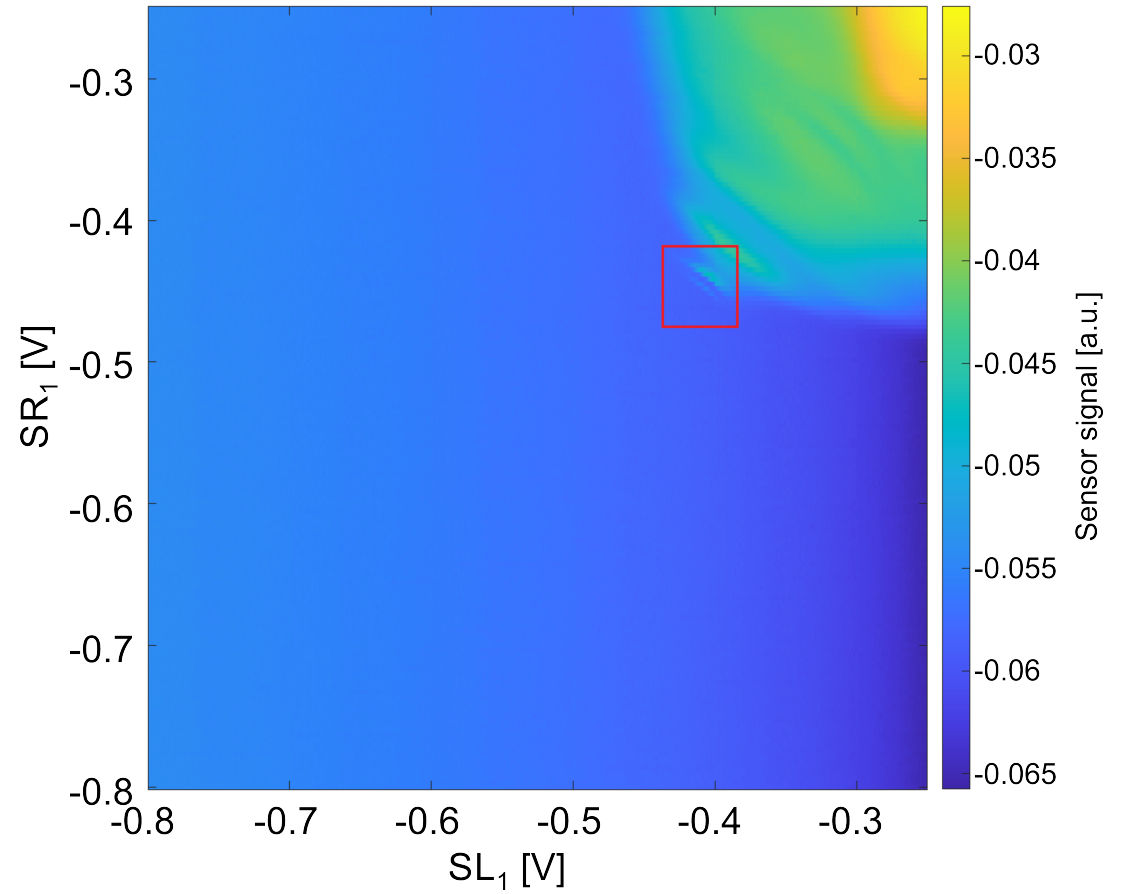}
    	\caption{Wide scan data of a \sd{} with a red box marking the area-of-interest that contains the Coulomb oscillations. In this area a \sd{} has been formed and can be further adjusted.}
    	\label{fig:sd_widescan}
    \end{figure}

    This work focuses on criteria to ultimately evaluate the eligibility of Coulomb peaks.
    Therefore, this analysis requires that the sensor dot has been formed, i.e. Coulomb oscillations\footnote{Coulomb oscillations are voltage-dependent oscillations in conductance through a quantum dot.} are visible in barrier-barrier scans.
    This requires numerous device-dependent tuning steps which for our device are:
    \begin{enumerate}
    	\item Deplete 2DEG below $DL$/$DR$ to split the sample in sensing and computation region.
    	\item \label{item:pickplunger} Pick a voltage for the plunger $SP_i$ of \sd{} \(i=1,2\) (see \cref{fig:sd_gates}) from a priori knowledge.
    	\item Test if Coulomb oscillations are visible in barrier-barrier (\(SL_i\)/\(SR_i\)) scans. If not, go back to \ref{item:pickplunger}. Otherwise, select the voltage range with Coulomb oscillations.
    \end{enumerate}
    The last step relies on wide scan measurements using the same measurement principle as described in \cref{sssec:qc_data} for the narrow scans. The exemplary scan in \cref{fig:sd_widescan} covers a wide voltage range to determine the area with Coulomb oscillations (red box). Promising approaches for its localization include the application of:
    \begin{itemize}
    	\item classical image processing methods comprising
    	\begin{itemize}
    		\item the Gabor filter \cite{baart_2016},
    		\item region growing, edge detection, K-means clustering with contouring, and template matching \cite{uppuluru_2021} and
    	\end{itemize}
    	\item object detection with deep learning \cite{lennon_2019, nguyen_2021, uppuluru_2021}.
    \end{itemize}
    After performing the above steps, the measurement of the narrow scan data described in \cref{sssec:qc_data} enables the analysis presented in this work.
    
    \subsubsection{\Sd{} narrow scans}
    \label{sssec:qc_data}
    
\begin{figure}
    	\includegraphics[width=\linewidth]{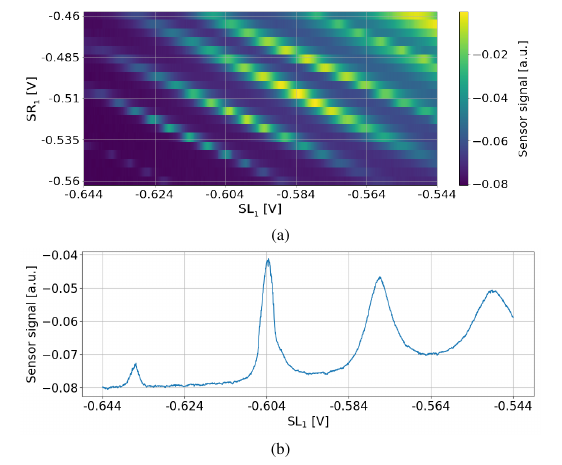}
    	\caption{Example data from sensor dot measurements. (a) \Sd{} narrow scan data. (b) Row data of the narrow scan in (a) with \(SR_1\) = -0.544 V.}
    	\label{fig:sd_data}
    \end{figure}

    Predominantly, the conductance characteristics of \sd{s} \(G_{sd}(SL, SR)\) represented in a \sd{} scan are controlled via the two barrier gate voltages (\(SL_i\) and \(SR_i\) for \sd{} \(i=1,2\) in \cref{fig:sd_gates}). A fine or narrow scan in the region with Coulomb oscillations (shown in \cref{fig:sd_data}a) sufficiently resolves the oscillations to determine an optimal sensor dot working point. These narrow scans are of interest to the present work. 
    
    Due to the experimental setup and measurement time reduction, the voltage on one axis changes step-wise. The other voltage continuously ramps up in a defined voltage interval, while an integration generates the measurement data. In our experiments, a narrow scan typically delivers 10 or 20 rows of data consisting of 1664 data points each. This distinct anisotropic 2D data resolution requires a row-wise data analysis to avoid artifacts. \cref{fig:sd_data}b shows an 1D data example of a single row. 
    
    \subsubsection{Requirements}
    \label{sssec:requirements_sdtun}
    
    To provide an optimal operating point, the requirements for a qualified region are:
    \begin{itemize}
    	\item a steep slope to achieve the strongest possible response and sensitivity\footnote{A steep slope as a function of the gate voltage corresponds to a steep slope as a function of the change of the electrostatic potential which is the parameter to be measured later on.} \cite{botzem_2017,zwolak_2021};
    	\item adequate linearity of the slope to achieve a constant behavior within the region\footnote{Automatic analysis can benefit from a linear response, e.g. for background compensation.};
    	\item a large region or rather a wide voltage range to have as much flexibility as possible when the working point shifts; and
    	\item a sufficient signal-gradient-to-noise ratio.
    \end{itemize}
    
\begin{figure}[!ht]
    	\centering
    	\captionsetup{singlelinecheck=off}
    	\includegraphics[width=0.9\linewidth]{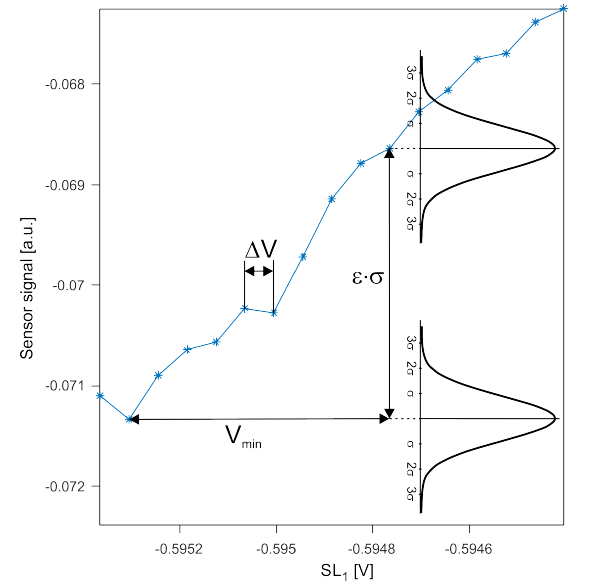}
    	\caption[Visualization of the \ac{msg} criterion]{Visualization of the \ac{msg} criterion (\cref{eqn:minimum_signal_gradient}) with
    		\begin{align*}
    			\epsilon{} & : \text{desired signal-gradient-to-noise factor,} \\
    			\sigma{} & : \text{estimated noise \ac{std},} \\
    			\Delta{}V & : \text{voltage sample distance, and} \\
    			V_{min} & : \text{minimal required voltage resolution of the sensor dot.}
    	\end{align*}}
    	\label{fig:msg_visualisation}
    \end{figure}

    The operation point selection is sensitive to the last requirement and, thus, requires a precise \nest{}. 
    Therefore, we propose the \ac{msg} criterion to ensure a sufficient signal-gradient-to-noise ratio in a qualified region:
    
    \begin{equation}
    	\label{eqn:minimum_signal_gradient}
    	MSG = \epsilon{} \cdot{} \sigma{} \cdot{} \frac{\Delta{}V}{V_{min}} \,,
    \end{equation}
    where \(\sigma{}\) denotes the \ac{std} of the estimated noise, \(\Delta{}V\) the voltage sample distance, \(V_{min}\) the minimal required voltage resolution of the \sd{}, and \(\epsilon{}\) the desired signal-gradient-to-noise factor. 
    \cref{fig:msg_visualisation} shows a visualization of the \ac{msg}'s composition. For a Gaussian distribution, 99.73\%{} of the noise is in the range of \(3\sigma{}\). We choose a value of 10 for \(\epsilon{}\), resulting in an additional security distance of \(4\sigma{}\) between two measurement points. 
    The term \(\frac{\Delta{}V}{V_{min}}\) relates the minimal required voltage resolution to sample points. 
    Thus, if the gradient in a qualified region is always greater than the \ac{msg}, the signal change for each \(V_{min}\) is at least \(\epsilon{}\) times as large as the noise \(\sigma{}\).
    
    \subsection{Noise estimation}
    \label{ssec:nest}
    
    Noise estimation is still a highly discussed topic in data analysis and processing. In signal processing, noise refers to unknown, unwanted distortions a signal may experience during acquisition, storage, transmission, processing, or conversion \cite{tuzlukov_2002}. From the estimation theory viewpoint \cite{kay_1993, chi_2006}, our 1D measurement (the observation) is regarded as one realisation of a sensor measurement (the random process). The parametric model described in \cref{ssec:noise_model} (the observation model) translates our a priori knowledge on the data. Primarily, our statistical signal part consists of Gaussian distributions (the probability laws) described by parameters. The task of the estimation is to find an approximate value for the parameters. In our case, only the estimation of the final noise amplitude's \ac{std} in the 1D data is required for the MSG criterion in \cref{eqn:minimum_signal_gradient}.
    
    However, existing state-of-the-art estimators often specialize in a particular application and data dimensionality. Therefore, to evaluate the estimators, we collect them, examine their adaptability to 1D space, and, if possible, revise them. The following subsections provide a classification overview on different noise estimation approaches. A more general insight to different noise estimators can be found in \cite {luo_2012, pandya_2014, liu_2013, liu_2014, khmag_2018}.
    
    \subsubsection{Spatial domain approaches}
    \label{sssec:spatial_dom_approaches}
    
    Approaches that work with spatial domain analysis apply:
    \begin{itemize}
    	\item local filtering techniques \cite{olsen_1993} and statistical analysis using
    	\begin{itemize}
    		\item the Laplacian filter \cite{immerkaer_1996},
    		\item Laplacian and gradient data masks \cite{corner_2003},
    		\item a modified Gaussian filter \cite{nguyen_2011}, and
    		\item the ROAD-statistic-based impulse detector \cite{garnett_2005},
    	\end{itemize} 
    	\item truncated local Taylor series approximation \cite{derrico_2007},
    	\item a step signal model utilizing polarized derivatives and their nonlinear combination \cite{laligant_2013},
    	\item a minimal controlled recursive averaging technique and neighborhood analysis \cite{mousavi_2016},
    	\item Mahalanobis distance measure from multivariate empirical mode decomposition (MEMD) \cite{rehman_2019},
    	\item a difference pre-filtering and statistical evaluation of a histogram of local signal variances \cite{rank_1999}, and
    	\item singular value decomposition (tail part) and content-dependent parameter determination \cite{liu_2013}.
    \end{itemize} 
    
    \subsubsection{Patch-based approaches}
    \label{sssec:patch_approaches}
    
    Patch-based approaches subdivide the data into homogeneous regions (patches, blocks, or superpixels). Patches are analyzed with different approaches using:
    \begin{itemize}
    	\item the mean of local variance of homogeneous regions \cite{bosco_2005},
    	\item a gradient approach (Laplace) \cite{tai_2008},
    	\item the Sobel filter and an averaging filter \cite{yang_2010},
    	\item Kendall‘s \(\tau{}\) \cite{schmidt_2011, sutour_2015a}, and
    	\item principal component analysis \cite{pyatykh_2014} applying for patch selection
    	\begin{itemize} 
    		\item static \cite{lee_1989, pyatykh_2013a} and adaptive \cite{shin_2005} thresholding or statistics of the patches' local variance \cite{rakhshanfar_2016},
    		\item a multi-directional high-pass operator as uniformity estimator followed by the calculation of a quantity threshold of homogeneity measure \cite{amer_2005},
    		\item the eigenvalues of the gradient covariance matrix of the patches \cite{liu_2012, liu_2013a},
    		\item the statistical relationship between the noise variance and the eigenvalues of the covariance matrix of patches \cite {chen_2015}, and
    		\item an iterative strategy to adaptively choose the optimal set of patches \cite {colom_2016, jiang_2016, fang_2019}.
    	\end{itemize}
    \end{itemize}
    
    \subsubsection{Transform-based approaches}
    \label{sssec:transform_approaches}
    
    The methods in this category apply the transformation of the given data with different approaches for noise identification:
    \begin{itemize}
    	\item wavelet decomposition \cite{donoho_1994, donoho_1995, abdelnour_2001},
    	\item trained moments and cumulative distribution functions of wavelet components \cite{destefano_2004},
    	\item goodness-of-fit test on dual-tree complex wavelet transform coefficients \cite{naveed_2017},
    	\item Morrison noise reduction method \cite{smith_2006}, and
    	\item discrete cosine transformation \cite{ponomarenko_2003a, lebrun_2015, ponomarenko_2017a, colom_2013}
    	\begin{itemize}
    		\item with analysis of variance and kurtosis or \cite{zoran_2009}
    		\item utilizing a thin-plate spline approximation \cite{garcia_2010}.
    	\end{itemize}
    \end{itemize}

    \section{Evaluation procedure}
    \label{sec:eval_procedure}
    
    \subsection{Noise analysis}
    \label{ssec:noise_analysis}
    
    To evaluate the estimators, we need to analyze the nature of the existing noise to construct simulated data as the ground truth. Random fluctuations and telegraph noise \cite{kafanov_2008} are not of interest here. Moreover, dedicated algorithms have to detect them and reject contaminated flanks.\\
    
\begin{figure}[!tbp]
    	\centering
    	\includegraphics[width=1\linewidth]{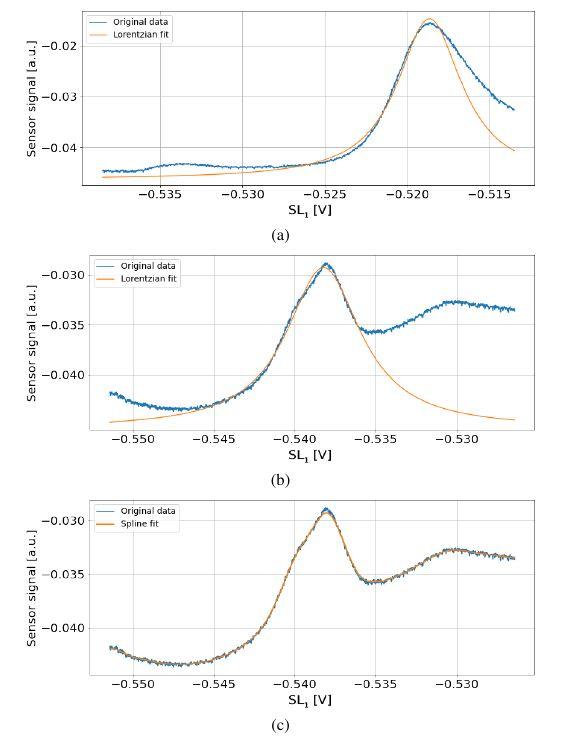}
    	\caption{Examples of problems when applying a fit with Lorentzians: (a) oscillations in the signal and missing information about the influence of the next peak and (b) unexpected peak shape and substantial interference between peaks. (c) The same data fit as in (b) but using Gaussian smoothing and spline interpolation.}
    	\label{fig:lorentzian_fit}
    \end{figure}

\begin{figure*}[htbp]
    	\centering
    	\includegraphics[width=1\linewidth]{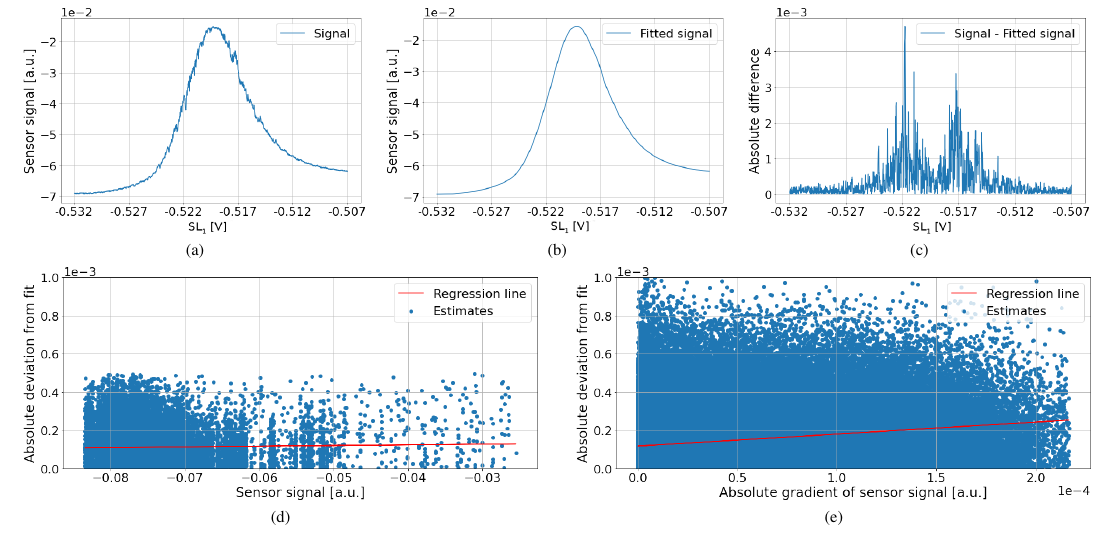}
    	\caption[Visual analysis of the noise using a fit]{Initial noise analysis by fitting the signal and calculating the difference in fit with the original signal. (a)-(c) Example showing (a) the signal, (b) the fitted version of the signal, and (c) the absolute difference between the signal and the fit. (d) Dependence of the noise on the sensor signal only in flat regions. Thus, we suppress the gradient-dependent parts. Signal peaks often cannot fulfill the flatness criterion, resulting in fewer estimates. The regression line shows a noise increase of 18.33\%{} and does not indicate any relevant dependence. (e) Dependence of the noise on the sensor signal gradient. The regression line shows a clear dependence on the gradient, with an increase of 115.52\%{}.}
    	\label{fig:visual_ana_region_dependency}
    \end{figure*}

    For noise analysis, a common strategy is to separate the noise from the signal by reconstructing the initial signal. However, various problems arise when fitting the signal with Lorentzians \cite{breit_1936}, the well-known behavior of Coulomb oscillations \cite{kouwenhoven_1997}. Usually, the fit works well for single peaks not influenced by neighboring peaks. However, peaks overlap in most of our data, challenging a suitable fit. Notably, the peak position is unknown for incomplete oscillations at the signal borders. We also observe several other distortions diverging from the Lorentzian model, including an overall signal drift, oscillations in the signal (see \cref{fig:lorentzian_fit}a), and peaks that do not conform to the expected shape (see \cref{fig:lorentzian_fit}b). Moreover, fitting with Lorentzians is considerably more time-consuming, as the dimensionality of the optimization problem increases with the number of peaks and sometimes requires manual adjustments.\\
    Therefore, we follow a general, fully automatic approach to extract the noise. First, we perform 1D Gaussian smoothing followed by a spline interpolation to ensure signal steadiness. Then, we subtract the approximated signal from the original signal to reveal a first assessment of the present noise, as visible in \cref{fig:visual_ana_region_dependency}. The example suggests that the noise characteristic is not homogeneous but depends on the signal regime.
    
\begin{figure}
    	\centering
    	\includegraphics[width=1\linewidth]{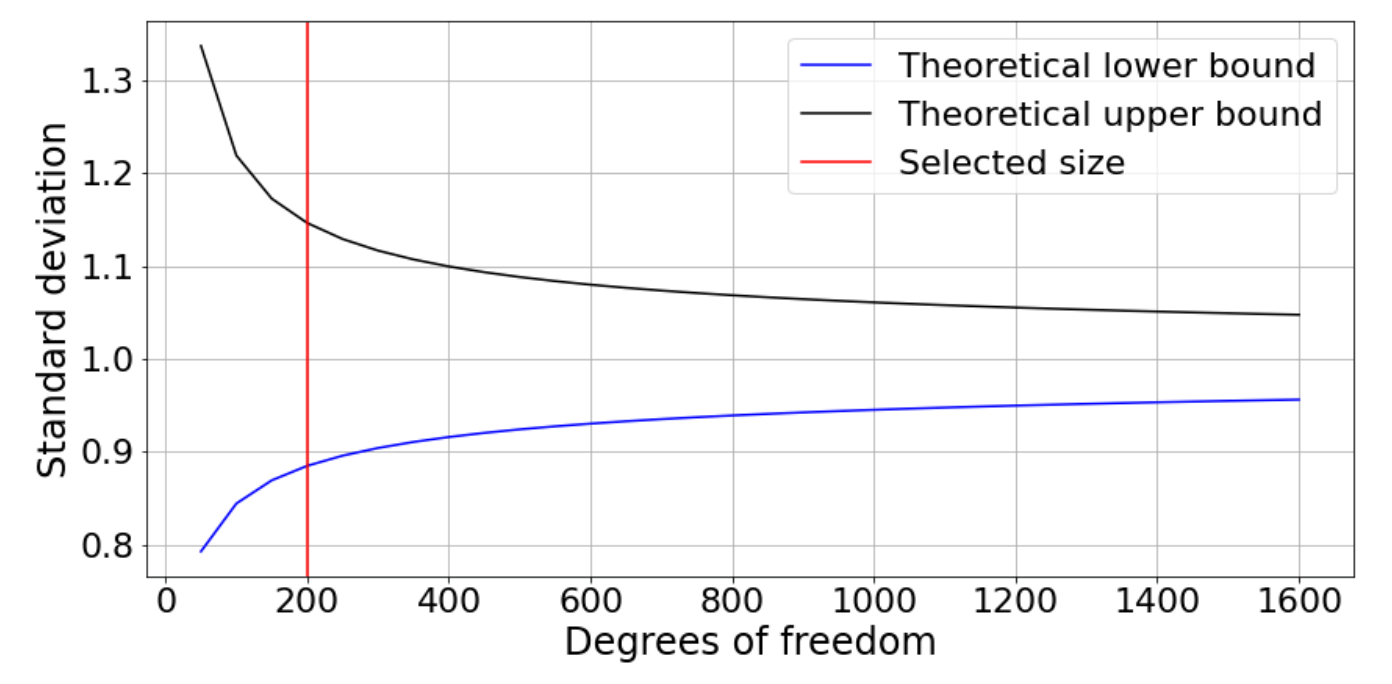}
    	\caption{The theoretical confidence interval for the \ac{std} of 1 with alpha=0.01 and different degrees of freedom.}
    	\label{fig:confidence_interval}
    \end{figure}

    To further investigate the signal dependence of the noise, we perform a local \nest{} using the methods of Chen et al.\footnote{With a patch size of 6} and Donoho \cite{chen_2015, donoho_1995} in a sliding window. Its size should balance sufficient statistics for a robust estimate with enough local significance. Therefore, we choose a window size of 201, considering the confidence intervals for the standard deviation of a draw from a normal distribution\footnote{Calculated using the chi-squared distribution} (see \cref{fig:confidence_interval}). Within the local noise analysis framework, we could confirm more substantial local noise in the flanks and a more distinctive overall noise scattering compared to the pure \ac{awgn} case. Among others, these originate from distortions directly affecting the gate voltages or occurring in the sample and are amplified parametrically by the sensor.
    
    \subsection{Noise model}
    \label{ssec:noise_model}
    
    \subsubsection{Standard model}
    \label{sssec:standard_model}
    
    The simplified generic model of the measured signal \(G_m\) consists of an idealistic noise-free signal \(G_{sd}\) at the sensor dot and additive noise \(n\), including all noise sources:
    \begin{equation}
    	\label{eqn:signal_model_generic}
    	\begin{aligned}
    		G_m(SL, SR) & = G_{sd}(SL, SR) + n \,.
    	\end{aligned}
    \end{equation} 
    \(n\) is composed of the \acl{ngrad} \acsu{ngrad} before and the \acl{nind} \acsu{nind} after the dot:
    \begin{equation}
    	\label{eqn:signal_model_generic_noise}
    	\begin{aligned}
    		n & = n_{grad} + n_{const} \\
    		& = X_{grad} \cdot{} \frac{\partial{}}{\partial{}SL} G_{sd}(SL, SR) + X_{const} \,,
    	\end{aligned}
    \end{equation} 
    with \(X_{grad}\) and \(X_{const}\) representing random variables for the noise distributions. According to \cref{ssec:tuning}, we presume pink noise for \ac{ngrad} and white noise for \ac{nind}. The amplitude of the individual noise sources is sufficiently independent and normally distributed. Based on the central limit theorem, we summarize both types of noise as one normal distribution. This results in the following representation:
    \begin{equation}
    	\label{eqn:noise_generic}
    	\begin{aligned}
    		X_{grad} & \sim \mathcal{N}(0,\,\sigma_{grad}^{2}) \\
    		X_{const} & \sim \mathcal{N}(0,\,\sigma_{const}^{2}) \,,
    	\end{aligned}
    \end{equation}
    with \(\sigma\) being the \ac{std} of the corresponding noise in the system. Because \(X_{grad}\) influences the applied voltages, it is multiplied by the signal gradient to obtain the measured noise \ac{ngrad}. In a sufficient linear region the corresponding \ac{std} \(\hat\sigma_{grad}\) can be calculated:
    \begin{equation}
    	\label{eqn:noise_model_sigma_grad}
    	\begin{aligned}
    		\hat\sigma_{grad} & = \sigma_{grad} \cdot{} \frac{\partial{}}{\partial{}SL} G_{sd}(SL, SR)\,.
    	\end{aligned}
    \end{equation}
    The \ac{std} of the total noise in the measurement data, \(\sigma_{n}\), is composed of \(\sigma_{const}\) and \(\hat\sigma_{grad}\):
    \begin{equation}
    	\label{eqn:noise_model_generic_sigma}
    	\begin{aligned}
    		\sigma_{n} & = \sqrt{\hat\sigma_{grad}^2 + \sigma_{const}^2} \,.
    	\end{aligned}
    \end{equation}
    
    \subsubsection{Extended model}
    \label{sssec:extended_model}
    
    Although we found no clear evidence for the presence of \acl{nsig} \acsu{nsig} in the measured data, we extended our model by \ac{nsig}, as it occurs in many applications:
    \begin{equation}
    	\label{eqn:noise_model_signal_generic}
    	\begin{aligned}
    		G_m(SL, SR) = & \; G_{sd}(SL, SR) + n_{grad}(SL, SR) \\ 
    		& + n_{const} + n_{lin}(SL, SR) \\
    		n_{lin}(SL, SR) = & \; X_{lin} \cdot{} G_{sd}(SL, SR)\\
    		X_{lin} \sim & \; \mathcal{N}(0,\,\sigma_{lin}^{2}) \\
    		\hat\sigma_{lin} = & \; \sigma_{lin} \cdot{} G_{sd}(SL, SR)\,.
    	\end{aligned}
    \end{equation} 
    We assume \ac{nsig} to be approximately white. The extended model allows for evaluating the estimators for a broader possible application spectrum.
    
    \subsection{Initial valuation for noise parameters}
    \label{ssec:initial_guess_noise}
    
\begin{figure}[!htbp]
    	\centering
    	\includegraphics[width=1\linewidth]{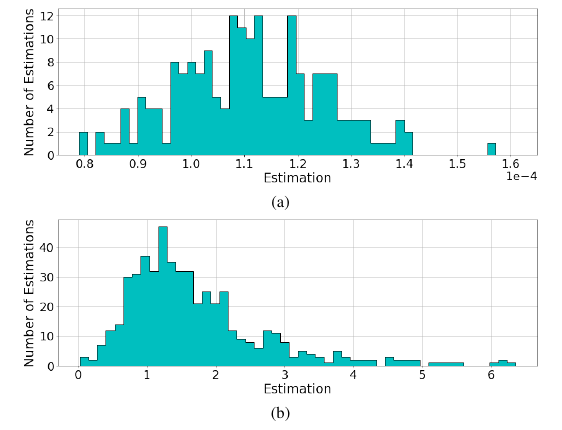}
    	\caption[Histogram of the initial guess for noise parameters]{Histogram of the estimated initial guess for the noise distributions \(\sigma{}\) obtained using the estimator of Donoho. (a) Estimates for \(\sigma{}_{const}\) and  (b) estimates for \(\sigma{}_{grad}\)}
    	\label{fig:initial_guess}
    \end{figure}

    We perform the first assessment for the different noise parts with the noise estimators of Donoho and Chen et al. Flat areas with a low slope in the one-dimensional measurement signal \(G_{m, 1D}(SL)\) qualify to determine an initial guess for \(\sigma_{const}\), whereas in steep areas, we estimate \(\sigma_{est}\). Then, we calculate an initial guess for \(\sigma_{grad}\):
    \begin{equation}
    	\label{eqn:initial_gradient}
    	\sigma_{grad} = \frac{\sqrt{\sigma_{est}^2-\sigma_{const}^2}}{\overline{\nabla{}G_{m, 1D}}} \,,
    \end{equation}
    where \(\overline{\nabla{}G_{m, 1D}}\) denotes the average gradient of an area. \cref{fig:initial_guess} illustrates the results for the estimator of Donoho. Furthermore, \cref{tab:results_initial_valuation} lists the corresponding \ac{cq95} interval for \(\sigma_{const}\), \(\sigma_{grad}\), and \(\hat\sigma_{grad}\). For the last, we multiply \(\sigma_{grad}\) with the average gradient of the dataset (\(\num{3.6e-5}\)). Notably, the \(\hat\sigma_{grad}\) interval scatters more than the \(\sigma_{const}\) interval, as the scattering of \(\sigma_{const}\) still affects the calculation of \(\sigma_{grad}\). The estimator of Chen et al. produces similar results but with less scattering.\\
    We also conduct an initial valuation for the linearly signal-dependent noise \ac{nsig}. For this purpose, we consider different levels of flat regions to suppress \ac{ngrad} and divide the estimates by the average signal of the corresponding region. Again, we cannot observe an existing \ac{nsig} of sufficient significance in these investigations.
    
    \begin{table}[!htbp]
    	\centering
    	\caption{Results of the initial valuation for different noise types using the estimator of Donoho.}
    	\label{tab:results_initial_valuation}
    	\begin{tabularx}{\linewidth}{X c}
    		\toprule
    		\textbf{Parameter} & \textbf{Value} \\
    		\midrule
    		\(CQ_{0.95}\) interval for \(\sigma_{const}\)  & \([8.6 ; 13.9] \cdot{} \num[print-unity-mantissa = false]{1e-5}\) \\
    		\midrule
    		\(CQ_{0.95}\) interval for \(\sigma_{grad}\) & \([0.4 ; 4.7]\) \\
    		\midrule
    		\(CQ_{0.95}\) interval for \(\hat\sigma_{grad}\) & \([1.4 ; 16.9] \cdot{} \num[print-unity-mantissa = false]{1e-5}\) \\
    		\bottomrule
    	\end{tabularx}
    \end{table}
    
    \subsection{Simulation of datasets}
    \label{ssec:eval_data}
    
    For our examination, we create two separate datasets. The first is to optimize the estimator parameters, and the second is to perform the quantitative analysis. The optimization dataset consists of 65 images and the evaluation dataset of 203 images. Both sets realize the standard and the extended noise model added to a fitted version of the measured signal. \cref{tab:stats_original_data} gives an overview of the value ranges of the data and gates.
    
    \begin{table*}
    	\centering
    	\caption{Parameters of the measured data. The gate voltages used differ between individual scans, and the individual dynamic voltage range falls into the interval of the respective gate. The stepsize indicates the difference in voltage between two measuring points.}
    	\label{tab:stats_original_data}
    	\begin{tabularx}{\linewidth}{Xccccccc}
    		\toprule
    		\textbf{Dataset} & \textbf{SL voltages} & \textbf{SL range} & \textbf{SL stepsize} & \textbf{SR voltages} & \textbf{SR range} & \textbf{Sensor signal} \\
    		& \textbf{[mV]} & \textbf{[mV]} & \textbf{[mV]} & \textbf{[mV]} & \textbf{[mV]} & \textbf{[\num[print-unity-mantissa = false]{1e-3} a.u.]} \\
    		\midrule
    		optimization & [-668.58, -398.03] & [25.00, 100.00] & [0.015, 0.060] & [-603.42, -442.69] & [25.00, 100.00] & [-84.85, 4.30] \\
    		\midrule
    		evaluation & [-694.74, -508.68] & [25.00, 100.00] & [0.015, 0.060] & [-691.31, -404.08] & [25.00, 100.00] & [-75.56, 12.01] \\
    		\bottomrule
    	\end{tabularx}
    \end{table*}
    
    Starting with the initial guess (see \cref{ssec:initial_guess_noise}), we further optimize the noise generation parameters for each image independently because of the scattering of their noise levels. The \ac{rmse} between the locally estimated \ac{std} of the measured and the simulated data steers optimizing the regionally differing noise in the complete signal. We choose the estimator of Chen et al. to target a low variation of the estimates\footnote{The data is not optimized for the estimator to estimate as accurately as possible, but to be as similar as possible to the real data. Thus, the estimator used has no advantage in the evaluation process by being used for data optimization. The underlying artificial noise and the accuracy of the estimation is not known to the estimator at any time}. A PCG family\footnote{PCG64 from \np{}-module \texttt{random}} generator, known for its excellent performance in statistical tests \cite{oneill_2014}, creates the \ac{awgn} part, whereas the algorithm from \cite{timmer_1995}\footnote{implemented in the python library \texttt{colorednoise}} produces the pink noise part.\\
    
\begin{figure}[htb]
    	\centering
    	\includegraphics[width=1\linewidth]{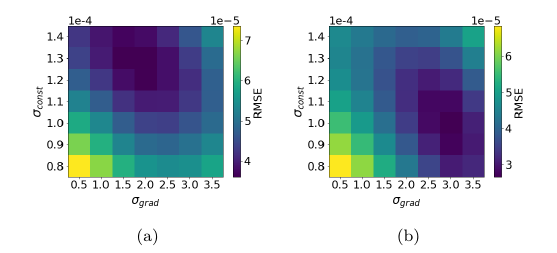}
    	\caption{Optimization results of noise parameters for the standard noise model. Heat map of the \ac{rmse} of noise parameter combinations across all images for (a) the optimization dataset and (b) the evaluation dataset.}
    	\label{fig:heatmap_combs_ind_w_grad_p}
    \end{figure}

    \cref{fig:heatmap_combs_ind_w_grad_p} shows the optimization results for standard noise model parameter combinations. We generate the simulated datasets with the best combination per image. 
    
    \subsection{Flank of interest detection}
    \label{ssec:func_find_region_candidates}
      
\begin{figure}[!tbp]
    	\centering
    	\includegraphics[width=1\linewidth]{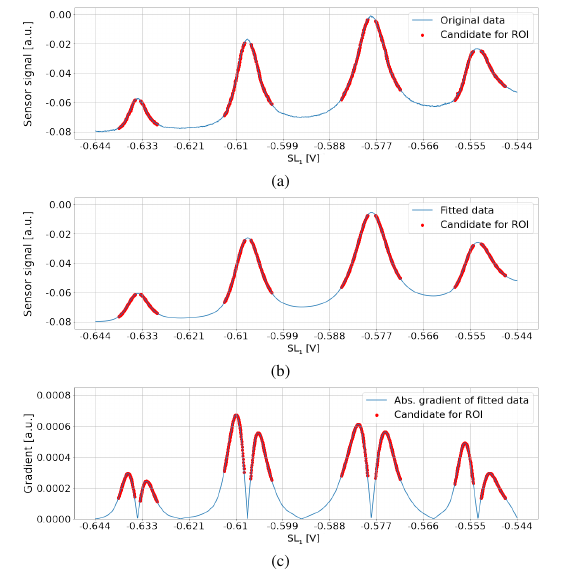}
    	\caption[\ac{roi} detection]{\ac{roi} detection: \ac{roi}s, defined by the minimal gradient, the minimal relative deviation from the local gradient maximum, and the minimal region size, are colored red. (a) A single row of measured data, (b) the fitted row, and (c) the absolute gradient of the fitted row.}
    	\label{fig:regionfinder_visualisation}
    \end{figure}

    First, our \ac{roi} detection algorithm calculates the absolute gradient of a fit to the measured data. Then, it selects maxima with a value of at least $discard\_weak$ times the value of the strongest one in the current row. Doing so dynamically prevents the inclusion of small maxima resulting from disturbances in the original signal. Next, around each selected maximum, the algorithm determines a candidate region defined by values higher than $increase\_until$ times the maximum. Finally, it records only candidate regions as \ac{roi}s with at least $min\_size$ data points determined from the sampling rate and a voltage range of \SI{3}{\mV}. Default values used for our \sd{} scans are:
    \begin{itemize}
    	\item $min\_size = \lceil{}\frac{\SI{3}{\mV}}{\Delta{}x}\rceil{}$
    	\item $increase\_until = 0.5$
    	\item $discard\_weak = 0.2$
    \end{itemize}
    The example in \cref{fig:regionfinder_visualisation} visualizes detected and selected edges in the signal.
    
    \subsection{Setup of estimators}
    \label{ssec:estors}
    
    \cref{tab:estimator_source_code} gives an overview of the estimators considered in our work, their references, and the source code used. We only consider estimators adaptable to 1D signals.
    
    \begin{table}[!htbp]
    	\centering
    	\small
    	\caption{Implementations of the noise estimators, sorted by the year of publication}
    	\label{tab:estimator_source_code}
    	\begin{tabularx}{\linewidth}{l X l l l} 
	    	\toprule
	    	\textbf{Estimator name} & \textbf{Year} & \textbf{Ref.} & \textbf{Source} & \textbf{Programming} \\
	    	& & & \textbf{Code} & \textbf{language} \\ 
	    	\midrule
	    	\textit{Olsen} & 1993 & \cite{olsen_1993} & \cite{laligant_2021} & Matlab \\
	    	\midrule
	    	\textit{DonohoJohnstone} & 1994 & \cite{donoho_1994} & \cite{foi_2011} & Matlab \\
	    	\midrule
	    	\textit{Donoho} & 1995 & \cite{donoho_1995} & \cite{foi_2011} & Matlab \\
	    	\midrule
	    	\textit{Immerk{\ae}r} & 1996 & \cite{immerkaer_1996} & \cite{foi_2011} & Matlab \\
	    	\midrule
	    	\textit{AbdelnourSelesnick} & 2000 & \cite{abdelnour_2001} & \cite{foi_2011} & Matlab \\
	    	\midrule
	    	\textit{Garnett} & 2005 & \cite{garnett_2005} & \cite{fleitmann_2022} & Matlab \\
	    	\midrule
	    	\textit{Smith} & 2006 & \cite{smith_2006} & \cite{smith_2006} & Matlab \\
	    	\midrule
	    	\textit{D'Errico} & 2007 & \cite{derrico_2007} & \cite{derrico_2007} & Matlab \\
	    	\midrule
	    	\textit{TaiYang} & 2008 & \cite{tai_2008} & \cite{laligant_2021} & Matlab \\
	    	\midrule
	    	\textit{ZoranWeiss} & 2009 & \cite{zoran_2009} & \cite{zoran_2009a} & Matlab \\
	    	\midrule
	    	\textit{Garcia} & 2010 & \cite{garcia_2010} & \cite{garcia_2020} & Matlab \\
	    	\midrule
	    	\textit{YangTai} & 2010 & \cite{yang_2010} & \cite{schwemmer_2019} & Matlab \\
	    	\midrule
	    	\textit{Liu} & 2012 & \cite{liu_2012} & \cite{tanaka_2015} & Matlab \\
	    	\midrule
	    	\textit{Laligant} & 2013 & \cite{laligant_2013} & \cite{laligant_2021} & Matlab \\
	    	\midrule
	    	\textit{Pyatykh} & 2013 & \cite{pyatykh_2013a} & \cite{pyatykh_2013} & Matlab \\
	    	\midrule
	    	\textit{Chen} & 2015 & \cite{chen_2015} & \cite{yue_2020} & Python \\
	    	\midrule
	    	\textit{Sutour} & 2015 & \cite{sutour_2015a} & \cite{sutour_2015} & Matlab \\
	    	\midrule
	    	\textit{Ponomarenko} & 2017 & \cite{ponomarenko_2017a} & \cite{ponomarenko_2017} & Matlab \\
	    	\bottomrule
	    \end{tabularx}
	\end{table}
    
    We use the simulated optimization dataset to optimize the parameters of the estimators. The cost function consists of the \ac{rmse} between the estimated and the actual \ac{std} of all detected \ac{roi}s. Instead of using the \ac{std} parameters of the noise generation, we calculate the actual \ac{std} of the added noise to improve the accuracy of the ground truth. \cref{tab:optimisation_results} shows the best parameters used for the evaluation. Since we found the same best parameters for both noise models, we omit a further comparison between the models at this point.\\
    Although initially working in manually selected regions, the \textit{YangTai} and \textit{Sutour} approaches are not applicable because their internal homogeneity criteria do not match the characteristics of our data. Therefore, we exclude these estimators from further evaluation. Due to similar reasons, the \textit{Liu} estimator only enters its first iteration in the estimation process. However, we maintain this approach because the initial estimation is competitive.
    
    \begin{table}[!htbp]
    	\centering
    	\small
    	\caption{Results of the noise estimator parameter optimization}
    	\label{tab:optimisation_results}
    	\begin{tabularx}{\linewidth}{lXr} 
	    	\toprule
	    	\textbf{Estimator name} & \textbf{Parameters} & \textbf{Value} \\
	    	\midrule
	    	\textit{Olsen} & Percentage of points having a low & 55.0 \\ 
	    	& gradient to take into account &  \\
	    	& \dotfill & \dotfill \\
	    	& Averaging type & 1.0 \\
	    	& (0='lms'; 1='mean') & \\
	    	\midrule
	    	\textit{Smith} & Maximum number of iterations to use in the Morrison denoising method & 1.0 \\ 
	    	\midrule
	    	\textit{TaiYang} & Threshold for excluding structures and details after edge detection & 100.0 \\
	    	\midrule
	    	\textit{ZoranWeiss} & Patch size & 4.0 \\
	    	\midrule
	    	\textit{YangTai} & Threshold for excluding structures and details after edge detection & 1.0 \\
	    	\midrule
	    	\textit{Liu} & Patch size & 2.0 \\
	    	\midrule
	    	\textit{Pyatykh} & Patch size & 6.0 \\ 
	    	\midrule
	    	\textit{Chen} & Patch size & 3.0 \\ 
	    	\midrule
	    	\textit{Ponomarenko} & Number of DCT coefficients of each image block to be used for further analysis & 4.0 \\ 
	    	\bottomrule
	    \end{tabularx}
	\end{table}
    
    \section{Results}
    \label{sec:results}
    
    \subsection{Quantitative analysis using simulated data}
    \label{ssec:results_artificial}
    
\begin{figure}[!tbp]
    	\centering
    	\includegraphics[width=1\linewidth]{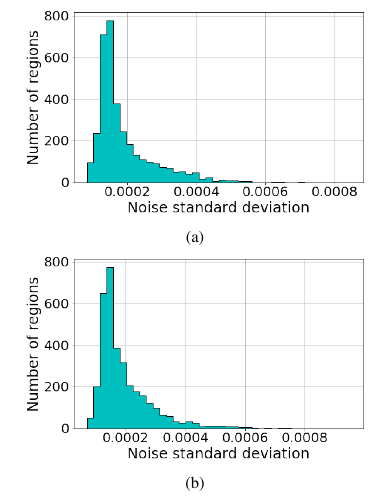}
    	\caption{Histogram of the actual noise \ac{std} in an \ac{roi} for the evaluation dataset: (a) standard noise model and (b) extended noise model.}
    	\label{fig:hist_true_stds_of_regions}
    \end{figure}

    We test the estimators on the evaluation dataset (see \cref{ssec:eval_data}) in regions detected with the algorithm described in \cref{ssec:func_find_region_candidates}. The ground truth is calculated the same way as in \cref{ssec:estors}. The histograms in \cref{fig:hist_true_stds_of_regions} show the distribution of \ac{roi}s to actual \ac{std}s for the two noise models. Considerably, both distributions are very similar, and most \ac{roi}s have an \ac{std} less than \(\num{2e-4}\).
    
\begin{figure*}[!htbp]
    	\centering
    	\includegraphics[width=1\linewidth]{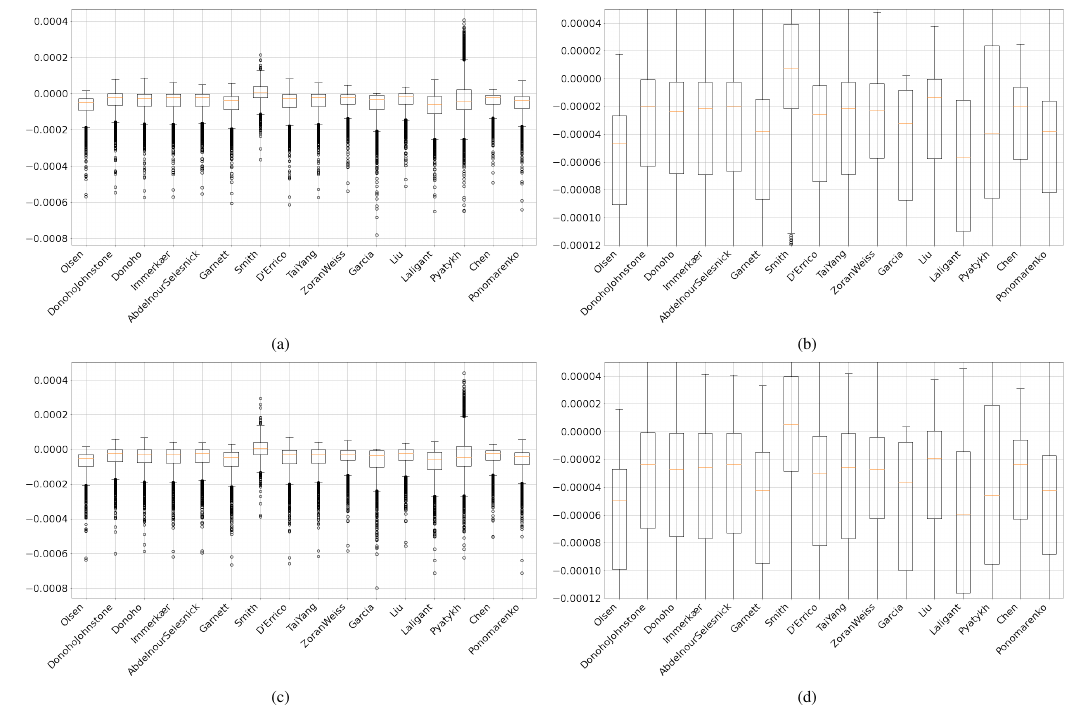}
    	\caption{Boxplots of the errors on the evaluation dataset. Each box contains the central 50\%{} of the data, bounded by the upper and lower quartiles. Its length corresponds to the \ac{iqr}, indicating the dispersion of the data. Furthermore, the red line in the box denotes the median. The whiskers represent the values outside the box, with lengths limited to 1.5 times the \ac{iqr}. All other values are treated as outliers and marked as dots. (a) Standard noise model, (b) Standard noise model clipped to the box, (c) extended noise model, and (d) extended noise model clipped to the box.}
    	\label{fig:boxplots_errors_finalset}
    \end{figure*}

    The boxplots in \cref{fig:boxplots_errors_finalset} show the \ac{std} estimation errors of the different approaches. The entire box of most estimators is negative, indicating their tendency to underestimate the noise. Additionally, notably, most estimators rarely show any upward outliers. However, the \textit{Pyatykh} and \textit{Smith} approaches have a symmetric distribution of errors around zero.\\
    In terms of the \ac{iqr} and the \ac{cq99r}, the best approaches are \textit{Chen}, \textit{ZoranWeiss}, \textit{Liu}, and \textit{Smith}. Overall, \textit{Chen} achieves the best results; only \textit{Smith} outperforms it on the extended noise model concerning the \ac{cq99r}.\\
    Similarly, these four estimators perform best concerning the \ac{rmse}. Here, the good result of \textit{Smith} derives from its mean close to zero and not from a good dispersion. Additionally, \cref{tab:quartiles_iqr_errors_ind_w_grad_p,tab:quartiles_iqr_errors_ind_w_grad_p_sig_w} in \cref{sec:appendix_supplementary_figures_tables} show the values for the quartiles, the associated \ac{iqr}, the \ac{cq99} interval, the \ac{cq99r}, and the \ac{rmse}.
    
\begin{figure}[!tbp]
    	\centering
    	\includegraphics[width=1\linewidth]{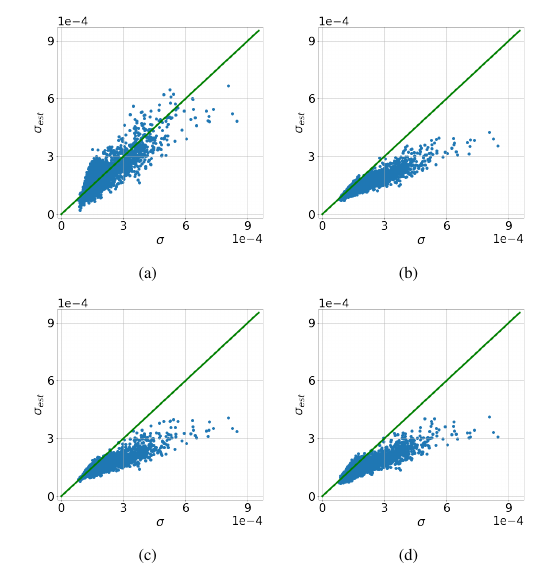}
    	\caption{Scatter plot (standard noise model) of the actual versus estimated \ac{std}s for the estimators of (a) \textit{Smith}, (b) \textit{Chen}, (c) \textit{Liu}, and (d) \textit{ZoranWeiss}. The green line in the figures references a perfect estimate. Each point in the scatter plot represents an estimation for one \ac{roi}. As expected, the estimates of the \textit{Smith} estimator scatter more evenly, whereas the other estimators tend to underestimate the noise. However, the \textit{Smith} method scatters more strongly than the other estimators.}
    	\label{fig:scatter_estimations_vs_real_stds_ind_w_grad_p}
    \end{figure}

    The scatter plots for the standard noise model in \cref{fig:scatter_estimations_vs_real_stds_ind_w_grad_p}\footnote{The plots of the other estimators as well as plots for the extended noise model are included in \cref{fig:scatter_standard_model,fig:scatter_extended_model} in \cref{sec:appendix_supplementary_figures_tables}} show the estimated and actual \ac{std} per \ac{roi} for the \textit{Smith}, \textit{Chen}, \textit{Liu}, and \textit{ZoranWeiss} estimators. The \textit{Chen} estimator has the lowest dispersion in compliance with the \ac{iqr} and \ac{cq99r} results. The underestimation of the noise is systematic and, therefore, can be approximately corrected with a compensation term. The \textit{Smith} estimator already uses such a correction term. Additionally, we include investigations on the performance of the estimators on pure \ac{awgn} and possible reasons for the underestimation behavior in \cref{sec:appendix_pure_awgn_estimations} and \cref{sec:appendix_investigation_underestimation}.
    
    \subsection{Qualitative evaluation: Detection of qualified regions in measured data }
    \label{ssec:results_real}
    
    Due to missing ground truth, a quantitative analysis of the estimation errors in the measured data is inapplicable. However, we can evaluate the impact of the estimators on the detection of qualified regions which comprise \ac{roi}s satisfying the \ac{msg} criterion from \cref{eqn:minimum_signal_gradient}.\\
    Therefore, we first determine \ac{roi}s using the algorithm described in \cref{ssec:func_find_region_candidates}. Then, for each \ac{roi}, we calculate the local \ac{msg} using an estimator to determine the \ac{std} of the local noise. Then, we determine the local gradient via a fitted version of the signal as in \cref{ssec:noise_analysis}. Finally, we check whether the absolute value of the local gradient consistently exceeds the \ac{msg}. If valid, we accept the \ac{roi} as a qualified region.
    
    \begin{figure}[!tbp]
    	\centering
    	\includegraphics[width=1\linewidth]{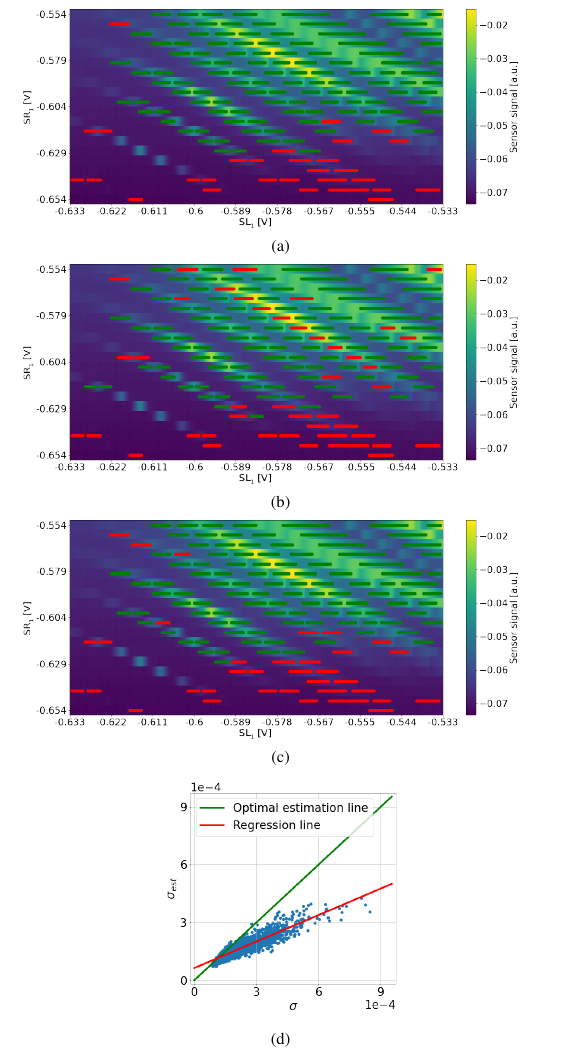}
    	\caption{Results of the investigation regarding the sensitivity of the quality of results to broader dispersion of the estimation: (a) \textit{Chen} approach, (b) \textit{Pyatykh} approach, and (c) \textit{Chen} approach with approximated compensation. Qualified regions are marked green, and rejected regions are marked red. (d) Regression line in \textit{Chen}'s scatter plot from \cref{fig:scatter_estimations_vs_real_stds_ind_w_grad_p}b.}
    	\label{fig:real_data_estimator_dispersion_investigation}
    \end{figure} 
    
    We perform the the qualified region detection with the \textit{Chen} (with and without underestimation compensation) and \textit{Pyatykh} estimator to examine their scatter and underestimation influence. Whereas the \textit{Chen} approach shows the lowest dispersion but an underestimation tendency, the \textit{Pyatykh} estimator features the highest dispersion.\\
    The results (\cref{fig:real_data_estimator_dispersion_investigation}) show that the flanks shift from row to row, forming diagonal wavefronts across several rows. Notably, rejected regions (red) are present where these wavefronts fade out because lower gradients no longer satisfy the \ac{msg} criterion. The above behavior is presumably less regular for an estimator with broader dispersion than with smaller dispersion.\\
    Accordingly, the \textit{Chen} estimator produces results without a random acceptance or rejection of \ac{roi}s (\cref{fig:real_data_estimator_dispersion_investigation}a). Moreover, the detection with the \textit{Pyatykh} estimator (\cref{fig:real_data_estimator_dispersion_investigation}b) shows rejections in strong and acceptance in low wavefront areas due to its higher scattering. A reduced signal-to-noise ratio due to the dominant constant noise floor intensifies this behavior in low wavefront areas. To assess the influence of the underestimation, we correct the \textit{Chen} estimator by an approximated compensation derived from the regression line in the examinations on simulated data (\cref{fig:real_data_estimator_dispersion_investigation}d). Again, the results (\cref{fig:real_data_estimator_dispersion_investigation}c) show no random behavior; only one might observe a negligible amount in the weak parts of the wavefronts.
    
    \section{Conclusion}
    \label{sec:conclusion}
    
    This article presents an approach to the noise-sensitive automated tuning of \sd{s} in gate-defined semiconductor quantum dots. First, we defined a minimal signal gradient criterion that requires a local noise estimate to identify qualified regions in \sd{} scans. Subsequently, our evaluation methodology for the estimators comprises a noise analysis, a noise model, the generation of simulated datasets for quantitative analysis, and an optimization of the estimator parameters. Then, the analysis based on the simulated data reveals the quantitative quality characteristics of the estimators and their tendency to underestimate. Finally, on measured data, we showed that a low estimation dispersion is of higher relevance for the region identification than the amount of underestimation.\\
    Therefore, when approximately compensating for the underestimation, we propose to use the estimator of Chen et al., as it performs best in terms of dispersion. Additionally, it only needs an average runtime of 0.5 milliseconds\footnote{Using an Intel Xeon E5-2630 v2 - 2.6 GHz} per estimated region and thus is not expected to be a limiting factor in an automated tuning process.
       
    Future research in the automation of the \sd{} tuning needs to address several aspects on basis of the results presented here. Dependent on the different application scenarios like tuning or readout, a flank selector has to optimize the balance between the flanks' width and steepness. Furthermore, the selector needs to map flanks to wavefronts, identify wavefront irregularities, and evaluate the neighborhood to select the most stable operation area. Additionally, an automated updating of the senor dot to compensate its drifts can significantly improve the quality of the readout data, and thus ease the automation of subsequent tuning steps. Eventually, all these steps have to trade off between complexity, flexibility and robustness to enable a hardware implementation that will be necessary for a fully functional quantum computer.
    
    \section{Data availability}
    \label{sec:data avail}
    
    The measured sensor dot data that support the findings of this study are available from Jülich DATA with the identifier \href{https://doi.org/10.26165/JUELICH-DATA/QIIBZV}{doi: 10.26165/JUELICH-DATA/QIIBZV}. All other data are available from the authors upon reasonable request.
    
    

    \ifCLASSOPTIONcaptionsoff
      \newpage
    \fi
    
    \bibliographystyle{IEEEtran}
    \bibliography{IEEEabrv,Literaturverzeichnis}

    \begin{IEEEbiography}[{\includegraphics[width=1in,height=1.25in,clip,keepaspectratio]{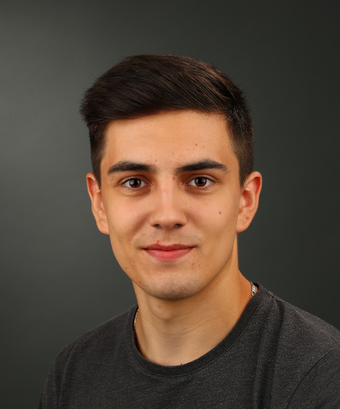}}]{Fabian Hader} received the B.Sc. degree in scientific programming and the M.Sc. degree in energy economics {\&} informatics from FH Aachen -  University of applied sciences, Jülich, Germany, in 2019 and 2021, respectively. He is currently pursuing a Ph.D. degree in engineering at University Duisburg-Essen, Duisburg/Essen, Germany.
    
    From 2019 to 2021, he was a Software Engineer at the Central Institute of Engineering, Electronics, and Analytics - Electronic Systems, Forschungszentrum Jülich GmbH, Jülich, Germany. His research interest focuses on the automatic tuning of quantum dots.
    \end{IEEEbiography}
    
    \begin{IEEEbiography}[{\includegraphics[width=1in,height=1.25in,clip,keepaspectratio]{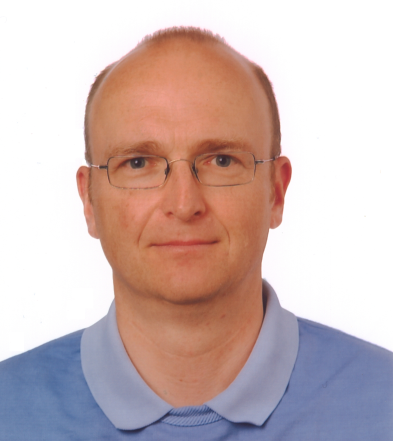}}]{Jan Vogelbruch} received the Dipl.Ing. and Dr.-Ing. degrees in electrical engineering from the RWTH Aachen University, Germany, in 1994 and 2003, respectively.
    
    In 1995, he joined Parsytec Computer GmbH, Aachen, Germany, as technical project manager for European cooperations. His focus has been on high-performance computing and image processing solutions, where he has been the technical leader for the company's part in several EC-funded projects.
    Since late 1998, he has been with the Central Institute of Engineering, Electronics, and Analytics - Electronic Systems at Forschungszentrum Jülich GmbH, Jülich, Germany. His research interests include parallel computing, signal and 3D image processing, fast reconstruction methods for high-resolution computer tomography, and automated defect detection. His current research focus is on the automatic tuning of semiconductor quantum dots.
    \end{IEEEbiography}

    \begin{IEEEbiography}[{\includegraphics[width=1in,height=1.25in,clip,keepaspectratio]{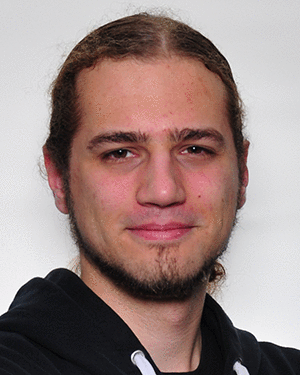}}]{Simon Humpohl} received the B.Sc. and M.Sc. degrees in physics in 2014 and 2017, respectively, from RWTH Aachen University, Aachen, Germany, where he is currently working toward the doctoral degree under the supervision of Prof. Hendrik Bluhm at the JARA Institute for Quantum Information.
    
    His research focuses on the tuning and operation of electron spin qubits in semiconductor quantum dots.
    \end{IEEEbiography}

    \begin{IEEEbiography}{Tobias Hangleiter} received the B.Sc. and M.Sc. degrees in physics in 2016 and 2019, respectively, from RWTH Aachen University, Aachen, Germany, where he is currently working toward the doctoral degree in quantum technology with 2nd Institute of Physics.
    
    His research interests include the automatic tuning of semiconductor quantum dots, quantum dynamics in the presence of correlated noise, and optical interfaces for semiconductor spin qubits.
    \end{IEEEbiography}

    \begin{IEEEbiography}[{\includegraphics[width=1in,height=1.25in,clip,keepaspectratio]{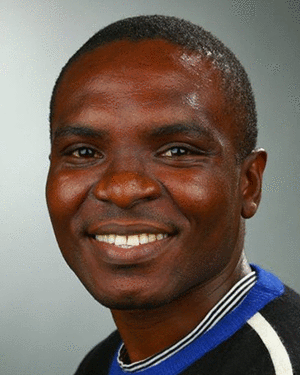}}]{Chimezie Eguzo} received the M.Sc. degree in microelectronics and wireless systems from Coventry University, Coventry, U.K., in 2014.
    
    In 2019, he joined the Central Institute of Engineering, Electronics and Analytics ZEA-2 Electronic Systems, Forschungszentrum Jülich, Jülich, Germany, where he is currently working on information and embedded systems. His research focuses on automating the build process and configuring the cross-compilation environment for different processing system architectures.
    \end{IEEEbiography}

    \begin{IEEEbiography}[{\includegraphics[width=1in,height=1.25in,clip,keepaspectratio]{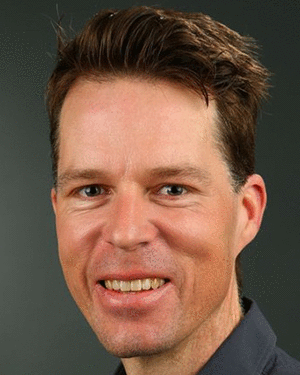}}]{Stefan Heinen} Stefan Heinen received the diploma and doctoral degrees in electrical engineering from the Aachen University of Technology, Aachen, Germany, in 1994 and 2001, respectively.
    
    In 2000, he joined Synopsys, working on the development of digital receiver algorithms for 2.5G, 3G, and GPS. His focus was on the algorithmic exploration and modeling methodology of HW/SW architectures. In 2004, he was a Virtual Prototype Architect with Infineon's mobile phone division, defining the model-driven design flow of Infineon's 3G/3.5G cell phone modem that later became part of the first iPhones. After the merger of Infineon's mobile business with Intel in 2011, he contributed to the definition of channel decoding and receiver algorithms for Intel's 4G and 5G cell phone modems. Since 2020, he has been with Helmholtz Research Center, Jülich, Germany. His research interests include speech and channel coding, error-robust parameter estimation, and digital signal processing algorithms applied to various fields, such as RF-based object localization and quantum computing.
    \end{IEEEbiography}
    
    \begin{IEEEbiography}[{\includegraphics[width=1in,height=1.25in,clip,keepaspectratio]{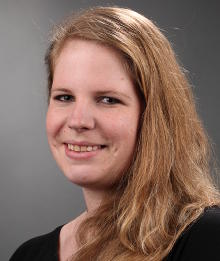}}]{Stefanie Meyer} received the B.Sc. degree in scientific programming and the M.Sc. degree in technomathematics from FH Aachen—University of Applied Sciences, Jülich, Germany, in 2011 and 2013, respectively.
    
    In 2008, she joined the Institute of Energy and Climate Research – Fuel Cells, Forschungszentrum Jülich GmbH, Jülich, Germany. Her research included developing and parallelizing high-temperature polymer electrolyte fuel cell models for HPC.\\
    Since 2014, she has been the head of the software development team at the Central Institute of Engineering, Electronics, and Analytics — Electronic Systems, Forschungszentrum Jülich GmbH, Jülich, Germany. Her research interests have a particular focus on electronic systems for quantum computing.
    \end{IEEEbiography}
    
    \begin{IEEEbiography}[{\includegraphics[width=1in,height=1.25in,clip,keepaspectratio]{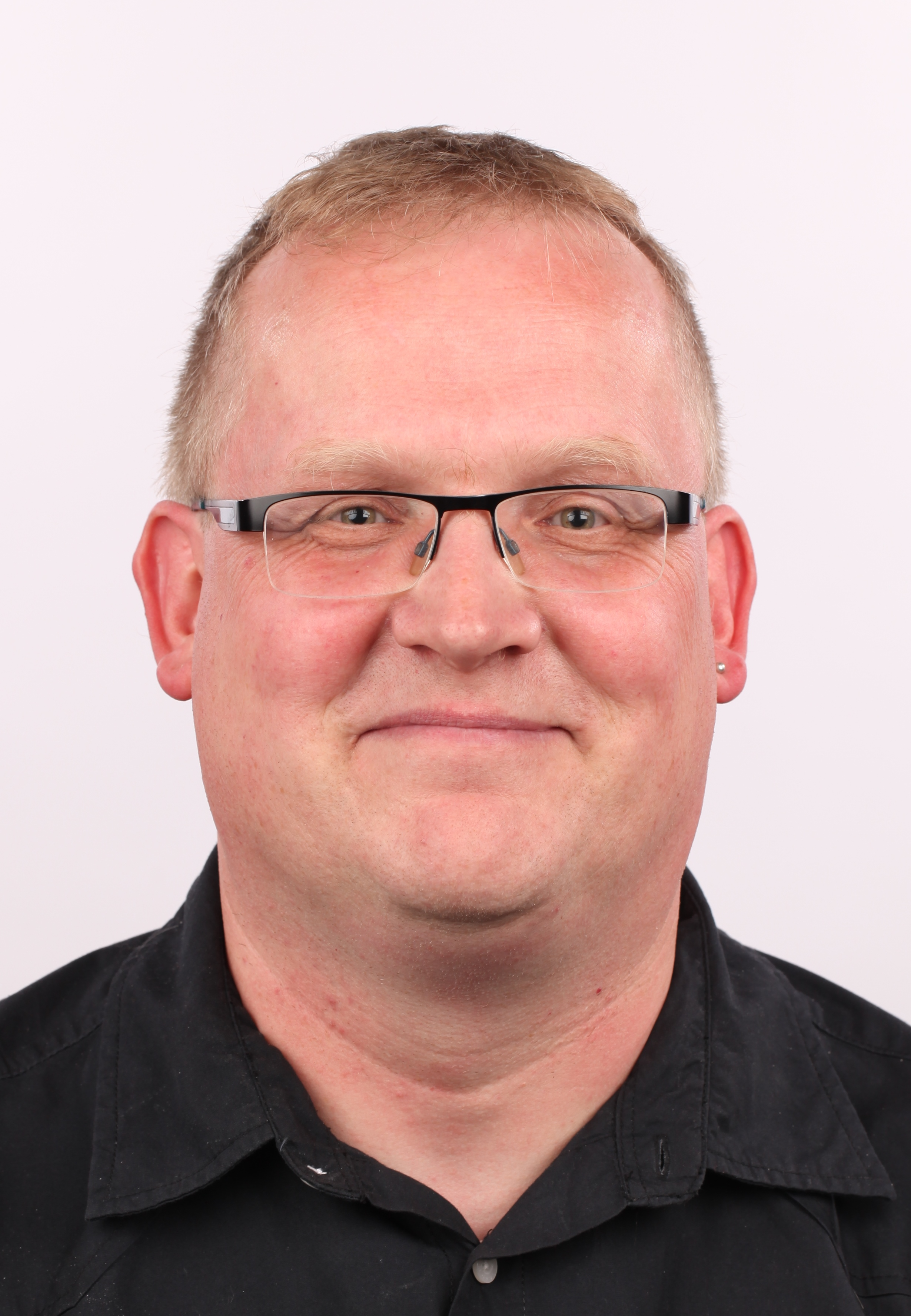}}]{Stefan van Waasen} received his diploma and doctor's degrees in electrical engineering from Gerhard-Mercator University, Duisburg, Germany, in 1994 and 1999, respectively. The topic of his doctoral thesis was optical receivers up to 60 Gb/s based on traveling wave amplifiers.
    
    In 1998, he joined Siemens Semiconductors/Infineon Technologies AG, Düsseldorf, Germany. His responsibility was BiCMOS and CMOS RF system development for highly integrated cordless systems like DECT and Bluetooth. In 2001, he changed into the IC development of front-end systems for high data rate optical communication systems. From 2004 to 2006, he was with the Stockholm Design Center responsible for the short-range analog, mixed-signal, and RF development for SoC CMOS solutions. From 2006 to 2010, he was responsible for the wireless RF system engineering in the area of SoC CMOS products at the headquarters in Munich, Germany, and later in the Design Center Duisburg, Duisburg. Since 2010, he has been the Director of the Central Institute of Engineering, Electronics, and Analytics - Electronic Systems at Forschungszentrum Jülich GmbH, Jülich, Germany. In 2014, he became a professor for measurement and sensor systems at the Communication Systems Chair of the University of Duisburg-Essen. His research is in the direction of complex measurement and detector systems, particularly on electronic systems for Quantum Computing.
    \end{IEEEbiography}
    
    \appendices
    \clearpage{}
    
    \setcounter{figure}{0} \renewcommand{\thefigure}{A.\arabic{figure}}
    \setcounter{table}{0} \renewcommand{\thetable}{A.\arabic{table}}
    
    \section{Behavior of estimators with pure AWGN}
    \label[appendix]{sec:appendix_pure_awgn_estimations}
    
\begin{figure}[!htbp]
    	\centering
		\includegraphics[width=1\linewidth]{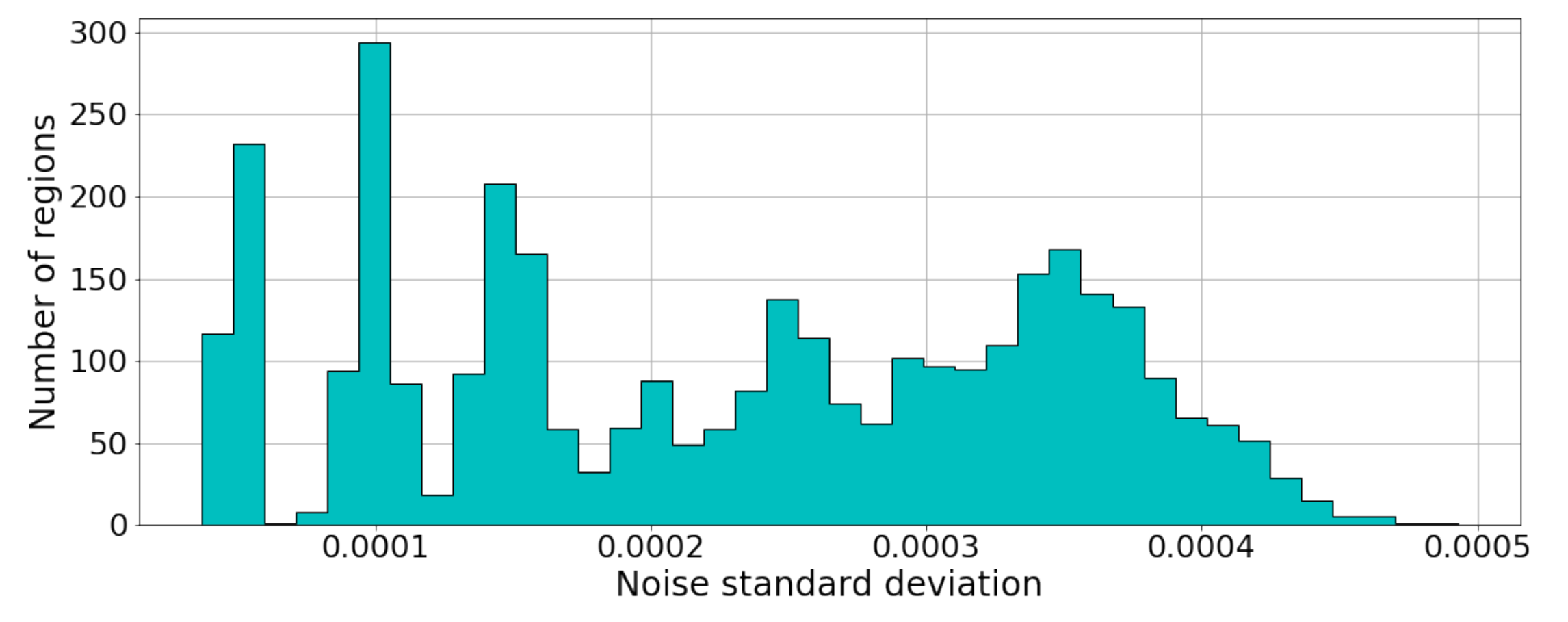}
    	\caption{Histogram of the distribution of the actual \ac{std}s in the \ac{roi} found on the evaluation dataset with pure \ac{awgn}}
    	\label{fig:hist_true_stds_of_regions_ind_w}
    \end{figure}

\begin{figure}[!htbp]
		\includegraphics[width=1\linewidth]{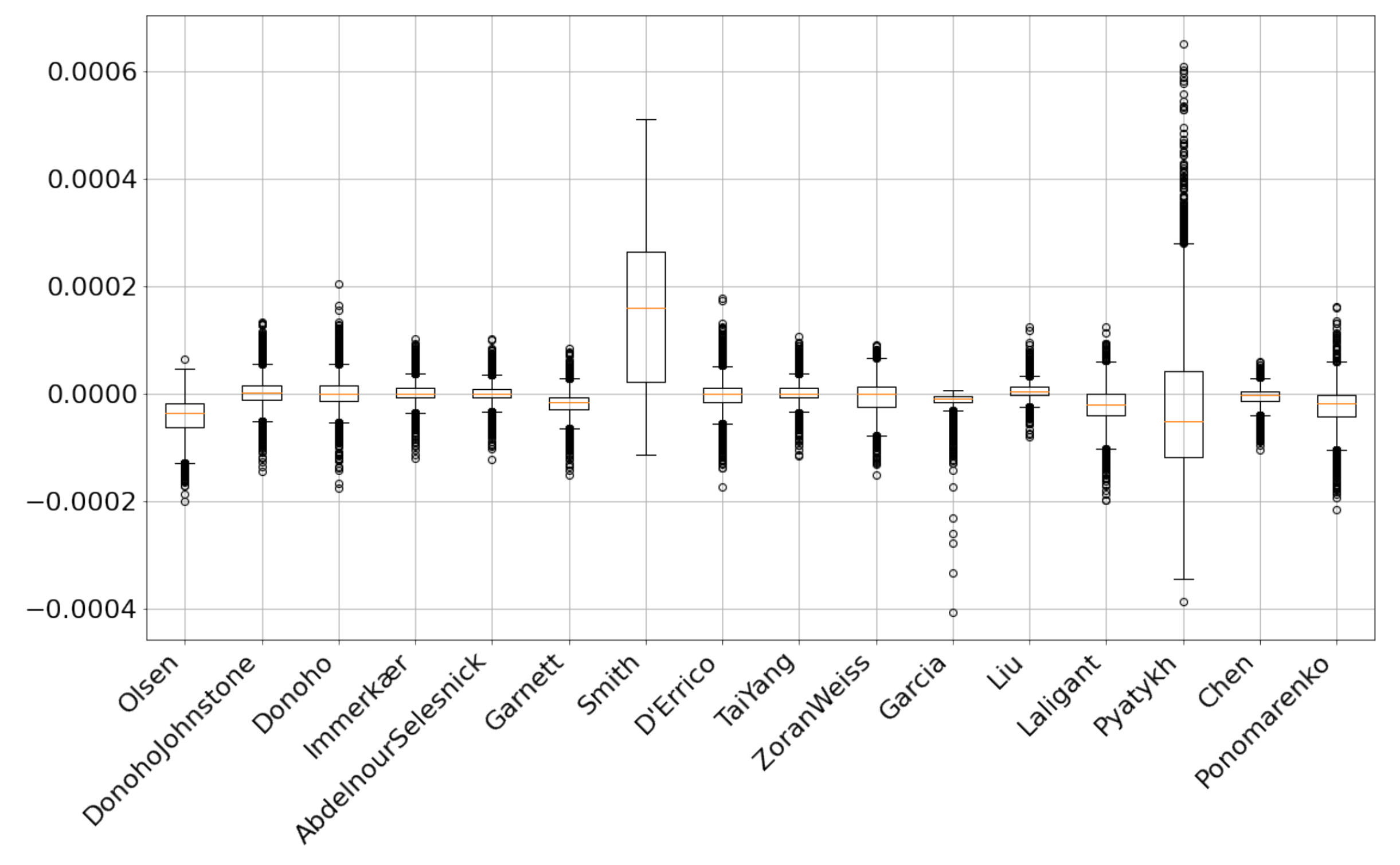}
    	\caption{Boxplot of errors on the evaluation dataset with pure \ac{awgn}}
    	\label{fig:boxplot_errors_finalset_ind_w}
    \end{figure}

    In addition to the tests with the developed noise model, we briefly evaluate the behavior of the estimators with pure \ac{awgn} to examine the influence of the type of noise on the estimation quality. Therefore, we add simulated noise with eight different \ac{std}s from the interval \([0.5, 4]\cdot{}\num[print-unity-mantissa = false]{1e-5}\) to the smoothed data. \cref{fig:hist_true_stds_of_regions_ind_w} shows the actual distribution of \ac{std}s in the \ac{roi}s, and \cref{fig:boxplot_errors_finalset_ind_w} shows the boxplot for the errors of the estimates. No systematic under- or overestimation of the noise occurs for most estimators.\\
    In conclusion, a systematic under- or overestimation strongly depends on the existing noise characteristics. Therefore, this must be considered when developing a compensation strategy to improve the estimation.
    
    \section{Sources for underestimation}
    \label[appendix]{sec:appendix_investigation_underestimation}
    
\begin{figure}[!htbp]
    	\captionsetup{singlelinecheck=off}
    	\centering
    	\includegraphics[width=1\linewidth]{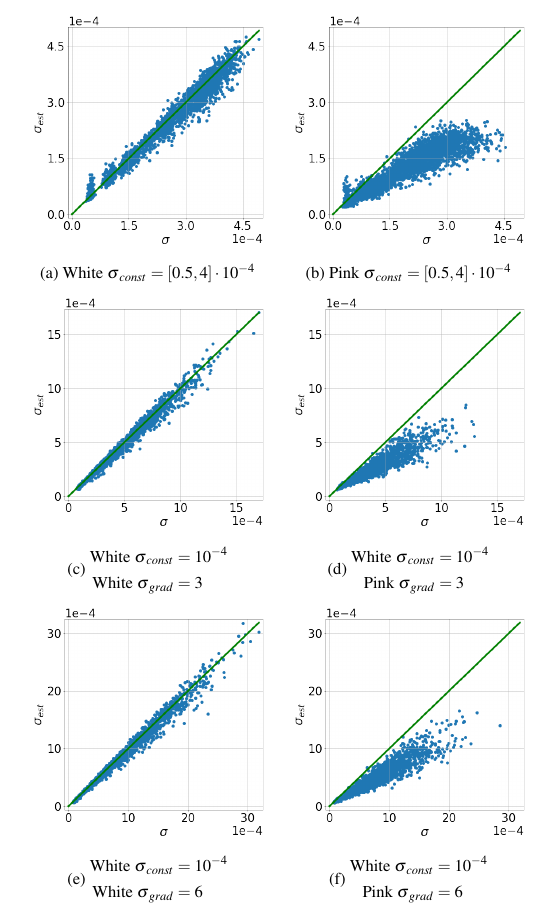}
    	\caption{Scatter plots of the actual versus the estimated \ac{std}s for different noise models. The underestimation does not depend on the regional dependence of the noise but rather on the presence of pink noise.}
    	\label{fig:investigation_underestimation}
    \end{figure}

    \cref{sec:appendix_pure_awgn_estimations} shows that underestimation does not occur with pure \ac{awgn}. Hence, the noise color and the regional dependence might cause the underestimation. Therefore, we apply different noise models to the evaluation dataset and exemplarily estimate the noise using the \textit{Chen} estimator. \cref{fig:investigation_underestimation} shows the results of the investigation. 
    
    Figs. \ref{fig:investigation_underestimation}a and \ref{fig:investigation_underestimation}b compare the results for independent \ac{awgn} and pink \ac{nind}. The following two subfigures show the results for a noise model consisting of white \ac{nind} and white \ac{ngrad} (\cref{fig:investigation_underestimation}c) and white \ac{nind} and pink \ac{ngrad} (\cref{fig:investigation_underestimation}d), respectively. Finally, Figs. \ref{fig:investigation_underestimation}e and \ref{fig:investigation_underestimation}f show results with a more intense gradient component. 
    In all three cases, underestimation only occurs when pink noise is present. In conclusion, the underestimation is not dependent on the regional dependence of the noise but only on the noise color. Accordingly, an underestimation compensation presumably requires knowledge of the predominant noise color. 
    
    \section{Supplementary figures and tables}
    \label[appendix]{sec:appendix_supplementary_figures_tables}
    
\begin{figure*}[!htbp]
    	\captionsetup{singlelinecheck=false,justification=raggedright}
    	\centering
    	\includegraphics[width=1\textwidth]{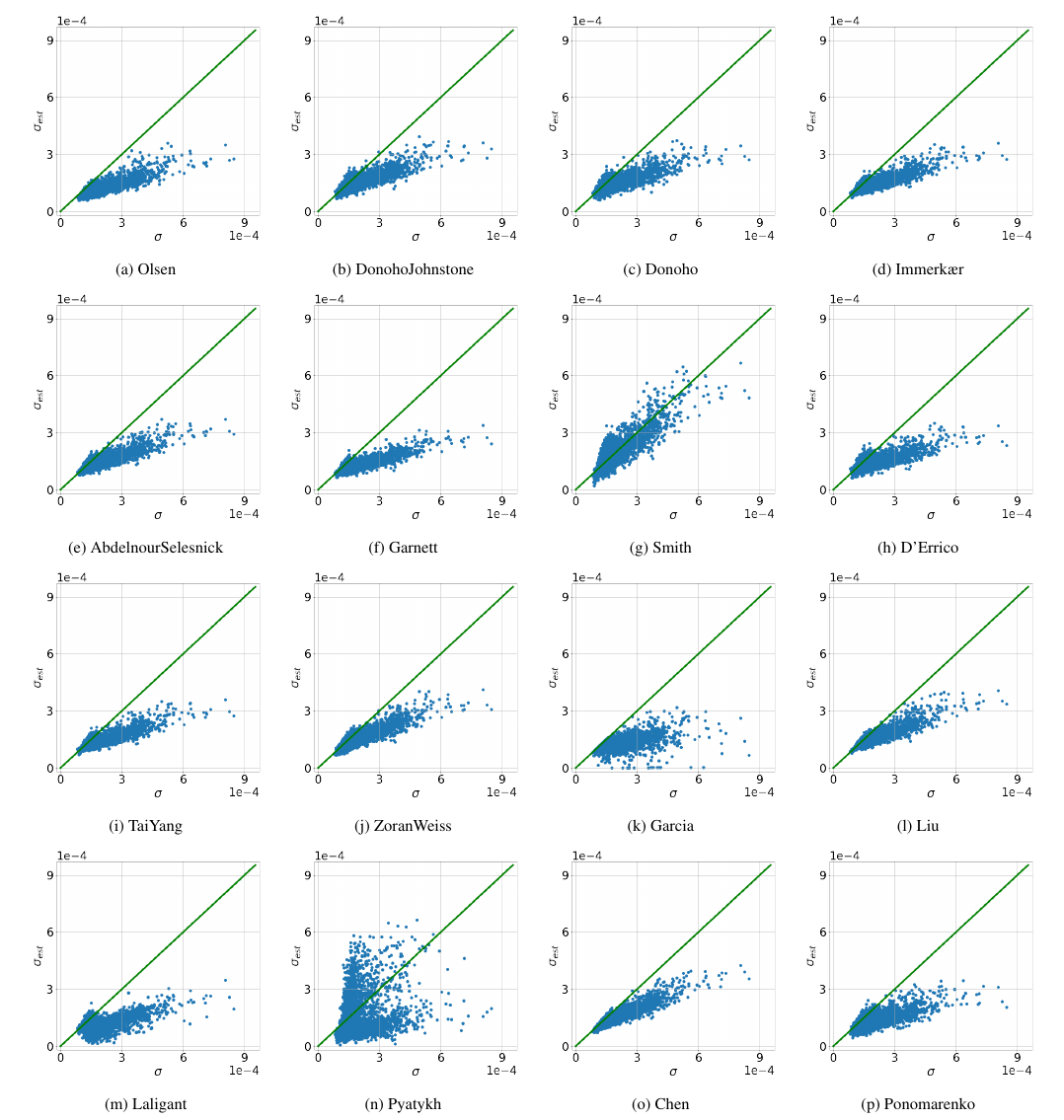}
    	\caption{Scatter plots of the actual versus the estimated \ac{std}s for the evaluation dataset with the standard noise model, sorted in the same order as in \cref{tab:estimator_source_code}. The green line in the figures indicates where the correct estimate should lie. Each point in the scatter plot represents an \ac{roi}.}
    	\label{fig:scatter_standard_model}
    \end{figure*}

\begin{figure*}[!htbp]
    	\captionsetup{singlelinecheck=false,justification=raggedright}
    	\centering
    	\includegraphics[width=1\textwidth]{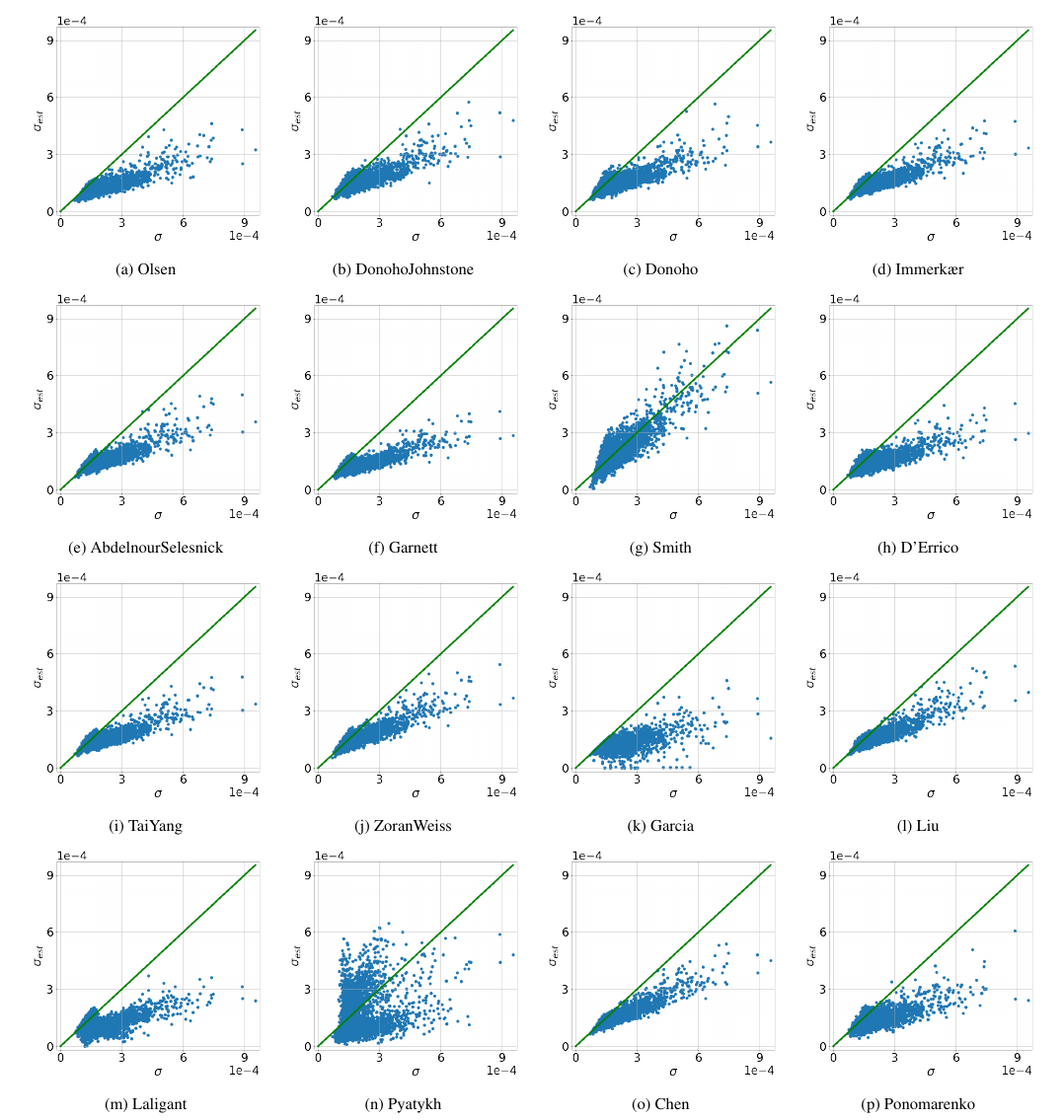}
    	\caption{Scatter plots of the actual versus the estimated \ac{std}s for the evaluation dataset with the extended noise model, sorted in the same order as in \cref{tab:estimator_source_code}. The green line in the figures indicates where the correct estimate should lie. Each point in the scatter plot represents an \ac{roi}.}
    	\label{fig:scatter_extended_model}
    \end{figure*}

    \cref{tab:quartiles_iqr_errors_ind_w_grad_p,tab:quartiles_iqr_errors_ind_w_grad_p_sig_w} show the values for the quartiles, the associated \ac{iqr}, the \ac{cq99} interval, the \ac{cq99r}, and the \ac{rmse} for the error of the estimates on the evaluation dataset. For all tested estimators, \cref{fig:scatter_standard_model,fig:scatter_extended_model} show the scatter plots of the actual versus the estimated \ac{std}s for the evaluation dataset.
    
    \FloatBarrier
    \begin{table*}[!htbp]
    	\centering
    	\caption{Quartiles, IQR, \(CQ_{0.99}\), \ac{cq99r}, and \ac{rmse} for the error of the estimates on the evaluation dataset with the standard noise model, sorted ascending by IQR}
    	\label{tab:quartiles_iqr_errors_ind_w_grad_p}
    	\begin{tabular}{lrrrrrrr}
    		\toprule
    		\textbf{Estimator} & \textbf{1st Quartile} & \textbf{3rd Quartile} & \textbf{IQR} & \(\mathbf{Q_{0.005}}\) & \(\mathbf{Q_{0.995}}\) & \(\mathbf{C_{0.99}R}\) & \textbf{RMSE} \\  
    		& \textbf{[\num[print-unity-mantissa = false]{1e-5}]} & \textbf{[\num[print-unity-mantissa = false]{1e-5}]} & \textbf{[\num[print-unity-mantissa = false]{1e-5}]} & \textbf{[\num[print-unity-mantissa = false]{1e-4}]} & \textbf{[\num[print-unity-mantissa = false]{1e-5}]} & \textbf{[\num[print-unity-mantissa = false]{1e-4}]} & \textbf{[\num[print-unity-mantissa = false]{1e-4}]} \\  
    		\midrule
    		\textit{Chen} & -5.79 & -0.63 & 5.16 & -2.60 & 1.39 & 2.73 & 0.65 \\
    		\midrule
    		\textit{ZoranWeiss} & -5.72 & -0.38 & 5.34 & -2.63 & 3.03 & 2.93 & 0.67 \\
    		\midrule
    		\textit{Liu} & -5.76 & -0.04 & 5.72 & -2.70 & 2.01 & 2.90 & 0.66 \\
    		\midrule
    		\textit{Smith} & -2.14 & 3.93 & 6.07 & -1.65 & 12.11 & 2.87 & 0.50 \\
    		\midrule
    		\textit{DonohoJohnstone} & -6.31 & -0.06 & 6.25 & -2.86 & 4.23 & 3.29 & 0.73 \\
    		\midrule
    		\textit{Olsen} & -9.07 & -2.68 & 6.39 & -3.26 & 0.18 & 3.28 & 0.93 \\
    		\midrule
    		\textit{AbdelnourSelesnick} & -6.66 & -0.24 & 6.42 & -2.93 & 2.73 & 3.20 & 0.75 \\
    		\midrule
    		\textit{Donoho} & -6.81 & -0.22 & 6.59 & -3.00 & 4.35 & 3.43 & 0.78 \\
    		\midrule
    		\textit{Ponomarenko} & -8.17 & -1.58 & 6.59 & -3.29 & 3.13 & 3.61 & 0.89 \\
    		\midrule
    		\textit{Immerk{\ae}r} & -6.88 & -0.28 & 6.61 & -2.96 & 2.74 & 3.23 & 0.77 \\
    		\midrule
    		\textit{TaiYang} & -6.87 & -0.26 & 6.61 & -2.95 & 2.82 & 3.24 & 0.77 \\
    		\midrule
    		\textit{D'Errico} & -7.40 & -0.49 & 6.91 & -3.04 & 4.47 & 3.49 & 0.82 \\
    		\midrule
    		\textit{Garnett} & -8.66 & -1.51 & 7.15 & -3.28 & 2.00 & 3.48 & 0.90 \\
    		\midrule
    		\textit{Garcia} & -8.76 & -0.82 & 7.93 & -4.09 & 0.10 & 4.10 & 1.01 \\
    		\midrule
    		\textit{Laligant} & -10.97 & -1.52 & 9.45 & -3.57 & 3.23 & 3.89 & 1.06 \\
    		\midrule
    		\textit{Pyatykh} & -8.62 & 2.36 & 10.98 & -3.81 & 31.62 & 6.97 & 1.24 \\
    		\bottomrule
    	\end{tabular}
    \end{table*}
        
    \begin{table*}[!htbp]
    	\centering
    	\caption{Quartiles, IQR, \(CQ_{0.99}\), \ac{cq99r}, and \ac{rmse} for the error of the estimates on the evaluation dataset with the extended noise model, sorted ascending by IQR}
    	\label{tab:quartiles_iqr_errors_ind_w_grad_p_sig_w}
    	\begin{tabular}{lrrrrrrr}
    		\toprule
    		\textbf{Estimator} & \textbf{1st Quartile} & \textbf{3rd Quartile} & \textbf{IQR} & \(\mathbf{Q_{0.005}}\) & \(\mathbf{Q_{0.995}}\) & \(\mathbf{C_{0.99}R}\) & \textbf{RMSE} \\  
    		& \textbf{[\num[print-unity-mantissa = false]{1e-5}]} & \textbf{[\num[print-unity-mantissa = false]{1e-5}]} & \textbf{[\num[print-unity-mantissa = false]{1e-5}]} & \textbf{[\num[print-unity-mantissa = false]{1e-4}]} & \textbf{[\num[print-unity-mantissa = false]{1e-5}]} & \textbf{[\num[print-unity-mantissa = false]{1e-4}]} & \textbf{[\num[print-unity-mantissa = false]{1e-4}]} \\ 
    		\midrule
    		\textit{Chen} & -6.28 & -0.58 & 5.70 & -2.91 & 1.75 & 3.08 & 0.69 \\
    		\midrule
    		\textit{ZoranWeiss} & -6.26 & -0.40 & 5.85 & -3.05 & 3.52 & 3.40 & 0.72 \\
    		\midrule
    		\textit{Liu} & -6.25 & 0.06 & 6.31 & -2.96 & 2.47 & 3.20 & 0.70 \\
    		\midrule
    		\textit{Smith} & -2.84 & 3.95 & 6.80 & -1.55 & 13.14 & 2.86 & 0.53 \\
    		\midrule
    		\textit{DonohoJohnstone} & -6.95 & -0.08 & 6.87 & -3.19 & 4.20 & 3.61 & 0.77 \\
    		\midrule
    		\textit{Ponomarenko} & -8.80 & -1.72 & 7.08 & -3.65 & 3.63 & 4.02 & 0.94 \\
    		\midrule
    		\textit{AbdelnourSelesnick} & -7.26 & -0.11 & 7.15 & -3.20 & 2.72 & 3.47 & 0.79 \\
    		\midrule
    		\textit{Olsen} & -9.88 & -2.69 & 7.19 & -3.64 & 0.63 & 3.70 & 0.98 \\
    		\midrule
    		\textit{Donoho} & -7.54 & -0.10 & 7.44 & -3.34 & 4.37 & 3.77 & 0.82 \\
    		\midrule
    		\textit{Immerk{\ae}r} & -7.66 & -0.14 & 7.52 & -3.34 & 2.92 & 3.63 & 0.82 \\
    		\midrule
    		\textit{TaiYang} & -7.67 & -0.13 & 7.53 & -3.33 & 3.05 & 3.64 & 0.82 \\
    		\midrule
    		\textit{D'Errico} & -8.19 & -0.34 & 7.85 & -3.60 & 3.87 & 3.99 & 0.88 \\
    		\midrule
    		\textit{Garnett} & -9.45 & -1.46 & 7.98 & -3.71 & 1.89 & 3.89 & 0.95 \\
    		\midrule
    		\textit{Garcia} & -9.97 & -0.74 & 9.22 & -4.58 & 0.12 & 4.59 & 1.08 \\
    		\midrule
    		\textit{Laligant} & -11.58 & -1.43 & 10.16 & -4.10 & 3.06 & 4.40 & 1.11 \\
    		\midrule
    		\textit{Pyatykh} & -9.51 & 1.89 & 11.41 & -4.13 & 31.23 & 7.25 & 1.24 \\
    		\bottomrule
    	\end{tabular}
    \end{table*}
    
    \FloatBarrier
    
    \FloatBarrier

    \FloatBarrier
    
    \EOD

\end{document}